\documentclass[aps,pra,twocolumn,nofootinbib,superscriptaddress,10pt]{revtex4-2}
\usepackage{amsmath,amsthm,amsfonts,amssymb,amscd,bbold,bm}
\usepackage[T1]{fontenc}
\usepackage{dcolumn}
\usepackage{physics}
\usepackage{graphicx}
\usepackage{xcolor}
\usepackage{epstopdf}
\usepackage{pgf}
\usepackage{tabularx}
\usepackage{subfigure}
\usepackage{xurl}
\usepackage[hidelinks]{hyperref}
\providecommand{\Eprint}[2]{}
\renewcommand{\Eprint}[2]{\href{#1}{\nolinkurl{#2}}}

\begin{document}

\title{Shaping of the finite-size chaos crossover in Floquet circuits by coherent-mismatch architecture}

\author{Xiangjun Tan}
\affiliation{Department of physics and astronomy, University College London, WC1E 6BT London, United Kingdom}
\author{Wenqi Liu}
\affiliation{School of Physical Science and Technology, Ningbo University, Ningbo 315211, China}
\author{Hanjie Zhu}
\affiliation{Center for Joint Quantum Studies and Department of Physics, School of Science, Tianjin University, Tianjin 300350, China}
\author{Wenkai Bai}
\email{baiw@nwu.edu.cn}
\affiliation{Shaanxi Key Laboratory for Theoretical Physics Frontiers, Institute of Modern Physics, Northwest University, Xi'an 710127, China}
\author{Zhanning Wang}
\email{z.w@csic.es}
\affiliation{Instituto de Ciencia de Materiales de Madrid, Consejo Superior de Investigaciones Cient\'ificas, 28049 Madrid, Spain}

\date{\today}
\begin{abstract}
We study how the spatial profile of a deterministic coherent gate perturbation affects finite-size quasienergy statistics and operator dynamics in $U(1)$-symmetric XXZ Floquet circuits.
Using a shared homogeneous reference, we compare a single modified bond with an alternating perturbation applied across the whole chain.
After separating symmetry sectors and checking several known integrable forms, we find that the alternating circuits move toward circular orthogonal ensemble level statistics and change operator-space entanglement and out-of-time-order correlations at smaller local amplitudes than the single defects.
For intermediate numbers of modified bonds, the response is nonmonotonic and also depends on where the bonds are placed, so bond density alone is not enough.
A perturbative analysis identifies the additive weight of many modified bonds, while coherent spatial interference reshapes the coupling between nearby eigenstates.
\end{abstract}
\maketitle

\section{Introduction}
\label{Sec1_introduction}
Integrability imposes constraints on many-body quantum dynamics, and breaking these constraints can lead to quantum chaos.
Understanding this change is central to nonequilibrium physics because it controls how information spreads and how isolated systems approach equilibrium.
This connection between spectral statistics, ergodicity, and thermalization is a central theme of quantum-chaos studies~\cite{Berry1977a,Bohigas1984,Rabson2004,Oganesyan2007,Atas2013,DAlessio2016,Santos2020}.

Floquet circuits provide a controlled setting for studying this crossover through a periodically repeated sequence of local quantum gates~\cite{DAlessio2014,Vanicat2018,Sieberer2019,Vernier2023,Miao2024}.
Their interactions, conservation laws, and spatial patterns can be tuned separately.
This platform links solvable many-body models to programmable quantum hardware.

Theory has uncovered several distinct routes between integrability and chaos in quantum circuits.
Random and generic circuits show how local interactions produce level repulsion, operator growth, and transport~\cite{Chan2018,Bertini2018,Kos2018,Nahum2018,vonKeyserlingk2018,Khemani2018,Rakovszky2018,Friedman2019,Bertini2020b,Alba2021,Winer2022}.
Special solvable circuits, however, show that rapid information spreading does not by itself imply chaos~\cite{Piroli2020,Bertini2021,Claeys2022,Gombor2022,Borsi2022,Miao2023,Fritzsch2025,Bertini2026}.
Homogeneous nearest-neighbor brickwall circuits of qubits that repeatedly apply the same $U(1)$-preserving two-qubit gate are integrable, including for generic choices of that gate~\cite{Duh2024,Znidaric2025a,Paletta2025a,Fernandez2026}.

At the opposite spatial extreme, one impurity can reshape the global many-body spectrum~\cite{Znidaric2020,SzaszSchagrin2021,Durnin2021,Surace2023}.
Many integrability-breaking gates can instead act as repeated local sources whose effects build up as information spreads~\cite{Kos2021,LopezPiqueres2022a,Riddell2024}.
Studies of decorated XXZ circuits further show that the number and regularity of added couplings can influence the dynamics~\cite{Hudomal2024,Paletta2025b}.
These studies identify three relevant features: the local gate, the number of modified bonds, and their spatial arrangement.

Recent experiments with quantum processors allow these features to be tested~\cite{Joshi2020,Nie2020,Mi2021,Blok2021,Braumuller2022,Green2022,Wang2022}.
Related platform studies have used random state preparation, spectral form factors, and dynamical observables to diagnose many-body chaos~\cite{Choi2023,Dong2025,Das2025,Fischer2026}.
Such processors can implement periodic XXZ circuits with tunable, charge-preserving gates and follow interacting excitations and correlations for many drive cycles~\cite{Chen2021,Zhao2022,Zhu2022,Nguyen2024,Shi2024,Zhang2024}.
Selected couplings can also move the circuit away from an integrable chain~\cite{Mi2022,Dong2023,Shtanko2025}.
A gate change can be applied to one bond, repeated across the device, or arranged in another fixed pattern.

The unresolved question is whether the response to the same local gate change depends only on how many bonds are modified or also on where they lie.
Most comparisons change the number of modified bonds, the overall perturbation, and the spatial pattern at once, making the role of placement hard to isolate.
Endpoint comparisons between one defect and full support also cannot reveal whether intermediate patterns follow a smooth density law.

Here, we show that the number of modified bonds is not a complete measure of integrability breaking.
We numerically study deterministic $U(1)$-preserving XXZ Floquet circuits while holding the local gate changes fixed.
Within each comparison between a single source and full support, we match the bond-averaged generator.
Applying the same charge-preserving gate deformation uniformly preserves integrability, whereas even one spatial mismatch can reshape the global spectrum.
A mismatch spread across the chain changes operator dynamics at a weaker local strength than a single defect.

In the largest circuits, full support also usually brings the spectral change to lower deformation amplitudes.
The path between one source and full support, however, is not smooth.
Even after two patterns contain the same signed sources in each brickwall layer and share the same spatial center, moving those sources changes the response and can reverse their ordering with size.

A bond-by-bond calculation explains why the endpoint separation is more stable.
We find that the accumulated weight of many local sources provides the leading effect, while interference redistributes which levels couple most strongly.
It does not consistently favor the nearest levels.
Source number sets the leading all-pair perturbative scale at fixed local amplitude, whereas placement controls finite-size shifts around it.

Section~\ref{Sec2_model} defines five circuit families and the expanded architecture tests, resolves their exact symmetries, and tests the nonuniform circuits against known integrable forms.
Section~\ref{Sec3_results} then tests the architecture dependence using short- and longer-range quasienergy statistics, operator entanglement, far-bond propagation, and local correlations across sizes and mismatch strengths.
Section~\ref{Sec4_perturbative} brings together the bond-resolved perturbative construction and its numerical evaluation to distinguish additive source contributions from interference between bonds.
Section~\ref{Sec3p6_architecture} then tests robustness across both XXZ parents, five generator pairs, every available odd source count, clustered and dispersed arrangements, time windows, and diagnostic thresholds.

\section{Model and theory}
\label{Sec2_model}
We consider a one-dimensional Floquet circuit of qubits with open boundary conditions and a conserved $U(1)$ charge.
We first define the integrable XXZ parent and decompose each local gate deformation into common and contrast generators.
We then use the contrast field to define the source count, placement, sign pattern, and homogeneous references for both the primary and expanded circuit families.
Finally, we inspect the exact symmetries and check known integrable structures.

\subsection{Floquet circuit and integrable XXZ chain}
\label{Sec2p1_floquet}
The open chain contains an even number $L$ of qubits labeled by $j$.
Pauli operators on each site are $X_j$, $Y_j$, and $Z_j$.
We define the local occupation operator $n_j$ and total charge $Q$ as:
\begin{equation}
n_j = \frac{\mathbb{1}+Z_j}{2}, \quad Q = \sum_{j=1}^Ln_j \,.
\end{equation}
Let $N$ denote an eigenvalue of $Q$, and let $\mathcal{H}_N$ be the corresponding fixed-charge sector.
We work at half filling $N=L/2$, with $d_N = \dim(\mathcal{H}_N) = \binom{L}{N}$.
We denote the projector onto this sector by $P_N$ and the identity operator on this sector by $\mathbb{1}_N$.

The open chain contains $M=L-1$ nearest-neighbor bonds.
Bond $b$ connects sites $(b,b+1)$ and carries the two-qubit gate $u_b$.
One Floquet period consists of two layers of non-overlapping gates:
\begin{equation}
U_1 = \prod_{\substack{1\leq b\leq M\\b\in2\mathbb{Z}+1}}u_b, \quad U_2 = \prod_{\substack{1\leq b\leq M\\b\in2\mathbb{Z}}}u_b \,.
\end{equation}
The one-period Floquet operator is $U_F = U_2U_1$.
Each local gate preserves the charge on its bond, so $[u_b,n_b+n_{b+1}]=0$ and $[U_F,Q]=0$.
The integer $t$ counts the number of completed Floquet periods, so stroboscopic evolution over $t$ periods is generated by $U_F^t$.
The Floquet eigenstates satisfy $U_F\ket{\alpha} = \mathrm{e}^{-\mathrm{i}\theta_\alpha}\ket{\alpha}$, where the eigenphases $\theta_\alpha$ are defined modulo $2\pi$.

The integrable parent uses the same charge-preserving XXZ gate on every bond.
Let $\Delta$ be the XXZ anisotropy.
The dimensionless two-qubit generator is:
\begin{equation}
h_{\text{XXZ}}(\Delta) = \frac{1}{4}[X\otimes X + Y\otimes Y + \Delta(Z\otimes Z - \mathbb{1}_4)] \,.
\end{equation}
For dimensionless gate duration $\tau$, the parent gate is $u_0(\Delta,\tau)=\mathrm{e}^{-\mathrm{i}\tau h_{\text{XXZ}}(\Delta)}$.
Setting $u_b=u_0(\Delta,\tau)$ on every bond gives the homogeneous integrable parent circuit.
Let $\delta\geq0$ denote the dimensionless local gate-deformation amplitude, with $\delta=0$ corresponding to the undeformed parent circuit.
We suppress the arguments $(\Delta,\tau)$ below and reserve the term parent for this gate and its $\delta=0$ circuit.

We fix $\tau=0.7$ and use the two parent anisotropies $\Delta=0.6$ and $\Delta=1.8$ on the gapless and gapped sides of the XXZ family, respectively.
At both values, placing the same gate on every bond gives an integrable XXZ Floquet circuit associated with the six-vertex Yang-Baxter construction~\cite{Ljubotina2019,Vernier2024,Hubner2025,Duh2026a}.

\subsection{Local generators and contrast field}
\label{Sec2p2_generators}
Our comparison requires two local gate deformations, whose common part can be held fixed while their difference is arranged across the chain.
We construct a pair of charge-preserving generators and decompose it into common and contrast components.

First, we define three diagonal directions as $Z_L=Z\otimes\mathbb{1}_2$, $Z_R=\mathbb{1}_2\otimes Z$, and $Z_LZ_R=Z\otimes Z$.
The convention $\sigma_\pm=(X\pm\mathrm{i}Y)/2$ is used.
The remaining two directions are the neutral hopping and current operators:
\begin{equation}
T = \sigma_+\otimes\sigma_- + \sigma_-\otimes\sigma_+, \quad J = \mathrm{i}(\sigma_+\otimes\sigma_- - \sigma_-\otimes\sigma_+) \,.
\end{equation}
All five operators commute with the two-qubit charge.

The primary generator pair, denoted by $G_0$, has nonzero components along all five traceless directions.
It represents a deterministic coherent difference between two gates, with no special physical significance assigned to the individual coefficient values.
The two unnormalized generators take the form $c_1 Z_L + c_2 Z_R + c_3 Z_LZ_R + c_4 T + c_5 J$, whose coefficients are $(0.37, -0.23, 0.41, 0.53, -0.47)$ and $(-0.29, 0.61, -0.35, 0.44, 0.57)$ for $W_{A,\text{raw}}$ and $W_{B,\text{raw}}$, respectively.
Tests with alternative generator pairs are reported in Sec.~\ref{Sec3p6_architecture} and Appendix~\ref{SecS5_architecture}.

For $\mu\in\{A,B\}$, we remove the identity component by defining $\tilde{W}_\mu=W_{\mu,\text{raw}}-(1/4)\Tr(W_{\mu,\text{raw}})\mathbb{1}_4$ and normalize it as $W_\mu=\tilde{W}_\mu/\|\tilde{W}_\mu\|_F$.
Here, $\norm{O}_F=\sqrt{\Tr(O^\dagger O)}$ denotes the Frobenius norm.
We omit the pair label on $W_A$ and $W_B$ when discussing the primary family.

The pair-average generator $W_{\text{avg}}$ is the common deformation, whereas the contrast generator $W_c$ changes sign between the two gates:
\begin{equation}
W_{\text{avg}} = \frac{W_A+W_B}{2}, \quad W_c = \frac{W_A-W_B}{2} \,.
\end{equation}
Because $W_A$ and $W_B$ have equal Frobenius norm, $\Tr(W_{\text{avg}}^\dagger W_c)=0$.
The three gates used below are $u_\mu(\delta)=\mathrm{e}^{-\mathrm{i}\delta W_\mu}u_0$ for $\mu\in\{A,B,\text{avg}\}$.

Each bond uses the common deformation plus a signed contrast source.
We encode the source on bond $b$ by a contrast field $s_b$ and write $W_b=W_{\text{avg}}+s_bW_c$ and $u_b(\delta)=\mathrm{e}^{-\mathrm{i}\delta W_b}u_0$.
The values $s_b=0$, $1$, and $-1$ select $W_{\text{avg}}$, $W_A$, and $W_B$, respectively.
An active bond or contrast source has $s_b\neq0$ relative to the homogeneous $u_{\text{avg}}$ background.
This distinction is relative to the homogeneous background because every gate can still differ from the parent gate when $\delta>0$.

\subsection{Contrast architectures}
\label{Sec2p3_architectures}
The complete signed field $\bm{s}=(s_1,\ldots,s_M)$ defines the contrast architecture.
Its active-bond set is $\mathcal{A}_{\bm{s}} = \{b\in\{1,\ldots,M\}:s_b\neq0\}$, which defines the source count as $K=|\mathcal{A}_{\bm{s}}|$ with source fraction $\rho=K/M$.
The signed bond average is:
\begin{equation}
\bar{s} = \frac{1}{M}\sum_{b=1}^Ms_b \,.
\end{equation}
The set $\mathcal{A}_{\bm{s}}$ specifies the bond placement, while the values of $s_b$ on that set specify the sign pattern.
The bond-averaged generator is:
\begin{equation}
\bar{W}_{\bm{s}} = \frac{1}{M}\sum_{b=1}^MW_b = W_{\text{avg}}+\bar{s}W_c \,.
\end{equation}

The five primary circuit families contain one homogeneous reference, two full-support alternating circuits, and two single-defect circuits.
The defects occupy adjacent central bonds, $b_A=L/2$ and $b_B=L/2-1$, which belong to opposite brickwall layers and share site $j=L/2$.
Their contrast fields are:
\begin{align}
\text{H}:& \quad s_b = 0 \\
\text{AB}:& \quad s_b = (-1)^{b-1} \\
\text{BA}:& \quad s_b = -(-1)^{b-1} \\
D_A:& \quad s_{b_A} = 1, \quad s_b = 0, \quad b\neq b_A \\
D_B:& \quad s_{b_B} = -1, \quad s_b = 0, \quad b\neq b_B \,.
\end{align}

The circuit $H$ is the shared homogeneous reference, whereas $AB$ and $BA$ are the two translations of the full-support alternating pattern.
The circuits $D_A$ and $D_B$ contain one contrast source in the same $u_{\text{avg}}$ background.
For fixed $L$, $N$, $\Delta$, $\tau$, and $\delta$, all five circuits share the parent gate, boundary condition, brickwall layout, and local generator pair.
They differ only through their architectures.

Because $L$ is even, $M=L-1$ is odd.
The values $(\bar{s},\rho)$ are $(0,0)$ for $H$, $(1/M,1)$ for $AB$, $(-1/M,1)$ for $BA$, $(1/M,1/M)$ for $D_A$, and $(-1/M,1/M)$ for $D_B$.
The pairs $AB$-$D_A$ and $BA$-$D_B$ therefore have the same bond-averaged generator and are called sign-matched pairs.
Each pair compares $K=M$ with $K=1$ at the same local amplitude $\delta$.

The total contrast weight $\sum_bs_b^2=K$ increases with the source count.
Generator matching is exact for the term linear in $\delta$, while nonlinear arithmetic gate averages can differ after exponentiation at finite $\delta$.

The expanded architecture family resolves the interval between the two source-count endpoints.
For every odd $K=1,3,\ldots,M$, we use a clustered active-bond set $\mathcal{A}_{K,\text{cl}}$ and a dispersed set $\mathcal{A}_{K,\text{disp}}$.
Their ordered bond indices are:
\begin{equation}
\begin{aligned}
b_{i,\text{cl}} &= \frac{M-K}{2}+1+i \\
b_{i,\text{disp}} &= 1+\left\lfloor\frac{i(M-1)}{K-1}+\frac{1}{2}\right\rfloor, \quad K>1 \,.
\end{aligned}
\end{equation}
For $K=1$, both placements use $b_0=(M+1)/2$, and for $K=M$, both sets contain every bond.

For placement $p\in\{\text{cl},\text{disp}\}$ and starting sign $\sigma\in\{1,-1\}$, the raw contrast field is:
\begin{equation}
s_{b,K,\sigma,p} = \begin{cases}\sigma(-1)^i, & b=b_{i,p}, \quad i=0,\ldots,K-1 \\ 0, & b\notin\mathcal{A}_{K,p}\end{cases} \,.
\end{equation}
The signs alternate by active-source rank, including across inactive gaps.
Because $K$ is odd, every placement satisfies:
\begin{equation}
\sum_{b=1}^Ms_{b,K,\sigma,p} = \sigma, \quad \bar{W}_{K,\sigma,p} = W_{\text{avg}}+\frac{\sigma}{M}W_c \,.
\end{equation}

\begin{figure*}[htbp!]
\centering
\includegraphics[width=\linewidth]{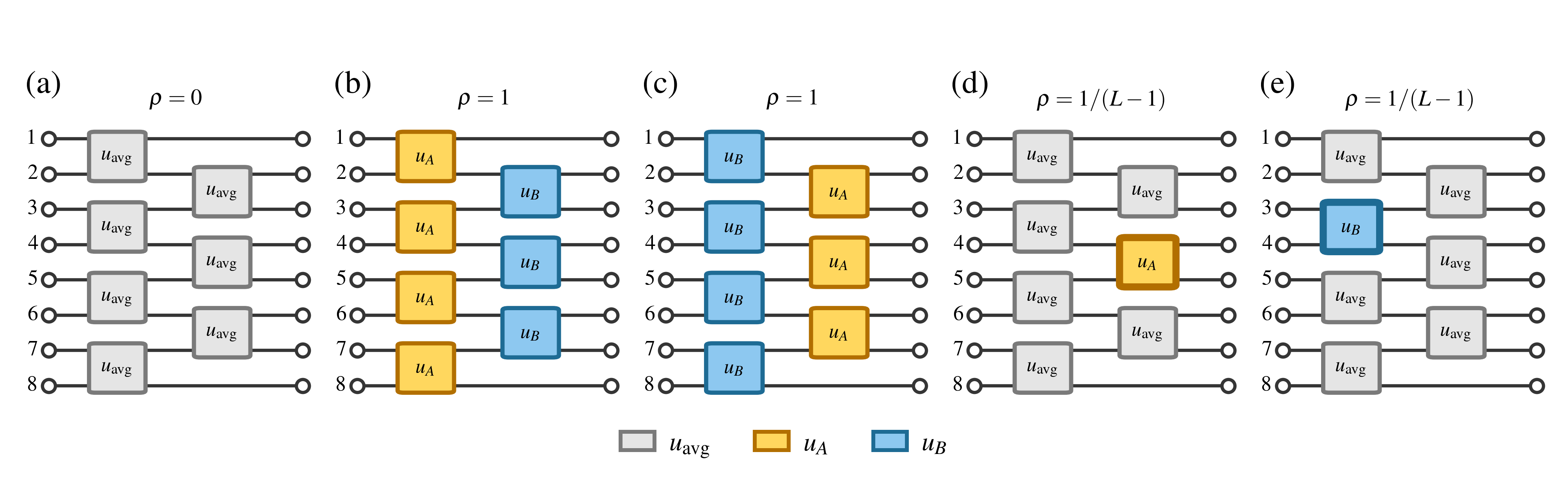}
\caption{Five gate patterns for one period of the open-boundary brickwall circuit, illustrated for $L=8$.
Time runs from left to right, with the first layer acting on odd bonds and the second layer acting on even bonds.
Grey, yellow, and blue denote $u_{\text{avg}}$, $u_A$, and $u_B$, respectively.
(a) The homogeneous reference $H$ uses $u_{\text{avg}}$ on every bond.
(b) and (c) The full-support circuits $AB$ and $BA$ alternate $u_A$ and $u_B$ with opposite starting gates.
(d) Circuit $D_A$ places $u_A$ on bond $b_A=L/2$ and uses $u_{\text{avg}}$ elsewhere.
(e) Circuit $D_B$ places $u_B$ on bond $b_B=L/2-1$ and uses $u_{\text{avg}}$ elsewhere.
The corresponding source fractions are $0$, $1$, $1$, $1/(L-1)$, and $1/(L-1)$.}
\label{Fig_1}
\end{figure*}

Each expanded architecture is compared with a sign-matched homogeneous control $H_\sigma$.
Its generator and local gate are $W_{\text{ctrl},\sigma}=W_{\text{avg}}+(\sigma/M)W_c$ and $u_{\text{ctrl},\sigma}(\delta)=\mathrm{e}^{-\mathrm{i}\delta W_{\text{ctrl},\sigma}}u_0$, respectively.
Relative to this control, the centered contrast field obeys $\tilde{s}_{b,K,\sigma,p}=s_{b,K,\sigma,p}-\sigma/M$, and the bond-generator difference is $W_{b,K,\sigma,p}-W_{\text{ctrl},\sigma}=\tilde{s}_{b,K,\sigma,p}W_c$.
Thus, $K$ counts the nonzero entries of the raw field $s_{b,K,\sigma,p}$, rather than all bonds whose generators differ from the homogeneous control.

The protocol fixes the local source amplitude and bond-averaged generator while allowing the source count, placement, total contrast weight, and spatial Fourier content to change.
The primary reference $H$ and the two expanded variations $H_\sigma$ form distinct comparison baselines.
We keep the nonuniform circuits fixed when comparing their responses to the different homogeneous references.

To isolate placement more strictly, we choose for each $L$ one intermediate odd source count nearest $M/2$, taking the smaller value when the two choices are equidistant.
We compare a compact field $s_{b,\text{comp}}$ with a dispersed field $s_{b,\text{disp}}$ after matching the source content within each brickwall layer.

Let $\Lambda_1$ and $\Lambda_2$ denote the odd- and even-bond layers.
The matched fields satisfy, for $q\in\{1,2\}$:
\begin{equation}
\hspace{-0.8cm}
\begin{aligned}
\sum_{b\in\Lambda_q}s_{b,\text{comp}}^2 &= \sum_{b\in\Lambda_q}s_{b,\text{disp}}^2, \quad \sum_{b\in\Lambda_q}s_{b,\text{comp}} = \sum_{b\in\Lambda_q}s_{b,\text{disp}} \\
\sum_{b=1}^M b s_{b,\text{comp}}^2 &= \sum_{b=1}^M b s_{b,\text{disp}}^2, \quad \sum_{b=1}^M b s_{b,\text{comp}} = \sum_{b=1}^M b s_{b,\text{disp}} \,.
\end{aligned}
\end{equation}
Thus the two fields have the same source count, signed generator mean, layer-by-layer source count and signed sum, active-support center, and signed first moment.
They differ only in higher spatial moments.
The negative-sign pair is the global sign reversal of the positive-sign pair (Appendix~\ref{SecS5_architecture}).

To test generator dependence, the expanded family uses $G_0$ and four additional fixed pairs $G_1$-$G_4$.
For $G_1$-$G_4$, independent standard-normal coefficient vectors are drawn in the Frobenius-orthonormal basis:
\begin{equation}
\mathcal{B} = \left(\frac{Z_L}{2},\frac{Z_R}{2},\frac{Z_LZ_R}{2},\frac{T}{\sqrt{2}},\frac{J}{\sqrt{2}}\right) \,.
\end{equation}
Each vector is normalized, and a pair is accepted only when $|\Tr(W_AW_B)|\leq0.65$ (Appendix~\ref{SecS5_architecture}).

\subsection{Symmetry resolution and random-matrix class}
\label{Sec2p4_symmetry}
Spectral statistics are meaningful only after independent symmetry sectors have been separated.
At $\delta=0$, every architecture and control reduces to the homogeneous XXZ parent.
In the half-filled sector, the parent has a global spin-flip symmetry $F$ and a spatial-reflection symmetry $R$:
\begin{equation}
F = \prod_{j=1}^LX_j, \quad R\ket{s_1s_2\cdots s_L} = \ket{s_Ls_{L-1}\cdots s_1} \,.
\end{equation}
Both operators commute with the parent Floquet operator $[U_F,F] = [U_F,R] = 0$.
The parent spectrum is resolved into simultaneous $(F,R)$ blocks with eigenvalues $f,r\in\{-1,1\}$.

For $\delta>0$, the chosen generator pairs break $F$, $R$, and their product $FR$ in the circuits used for spectral statistics.
We therefore analyze each positive-$\delta$ spectrum in the full fixed-$N$ sector rather than in the parent $(F,R)$ blocks.
These commutators are verified for every circuit and parameter set before the spectrum is analyzed.

The remaining symmetry determines which random-matrix ensemble provides the appropriate benchmark.
On an open chain, the phases of the charge-preserving hopping terms can be removed recursively by rotating the occupation basis at each site.
We write the corresponding unitary basis transformation as:
\begin{equation}
G = \bigotimes_{j=1}^L\mathrm{e}^{\mathrm{i}\phi_jn_j} \,.
\end{equation}
The transformed Floquet operator $\bar{U}_F=G^\dagger U_FG$ has the same quasienergies as $U_F$.
We denote its two layers by $\bar{U}_1$ and $\bar{U}_2$.

Shifting the start of the Floquet period to the middle of the first layer gives the symmetric time frame $U_{\text{sym}}=\bar{U}_1^{1/2}\bar{U}_2\bar{U}_1^{1/2}$, which satisfies $U_{\text{sym}}^T=U_{\text{sym}}$.
This symmetric-time-frame operator is unitarily equivalent to $\bar{U}_F$ and has the same quasienergies.
Complex conjugation in this basis defines an antiunitary operator $\mathcal{T}$ satisfying $\mathcal{T}U_{\text{sym}}\mathcal{T}^{-1}=U_{\text{sym}}^\dagger$.
This relation assigns the circular orthogonal ensemble (COE) symmetry class.
It determines the random-matrix benchmark but does not by itself imply COE spectral correlations.

\subsection{Checks for integrable structure}
\label{Sec2p5_integrability}

Spatial nonuniformity alone does not establish nonintegrability.
We test whether the four nonuniform members of the primary family fit several known integrable structures before interpreting their spectral statistics.
The circuit $H$ serves as the homogeneous reference and is distinct from the $\delta=0$ integrable parent.

To distinguish the Yang-Baxter matrix from the reflection operator $R$, we denote the former by $\mathsf{R}(\lambda)$, where $\lambda$ is its spectral parameter.
For the full-support circuits $AB$ and $BA$, we test whether both gates can come from one six-vertex $\mathsf{R}(\lambda)$ family after local diagonal changes of basis.
We also allow additional phases that depend only on the conserved two-site charge.
These embedding tests include direct and numerically optimized solutions of the braided Yang-Baxter relations.
The local basis phases are allowed to depend on the gate position and brickwall layer.

A further signature of many known integrable circuits is the presence of local conserved operators beyond the known charge.
We search successively larger operator-support spaces for nontrivial local operators $O$ satisfying $U_F^\dagger O U_F=O$.
The components generated by $\mathbb{1}$ and $Q$ are removed before the search.
For $AB$ and $BA$, the search respects the two-bond translation cell.
For $D_A$ and $D_B$, it uses real-space windows centered on the active bond.
We also test the local Yang-Baxter compatibility of each defect gate with the homogeneous background gate.

Within the tested parameter grids and operator-support cutoffs, none of the four nonuniform primary circuits admits a compatible embedding or an additional exact local conserved operator.
These negative tests rule out only the structures specified above within the searched ranges.

\section{Results and discussion}
\label{Sec3_results}
We first use symmetry-resolved spectral statistics to determine how full-support patterns and single defects move away from the integrable parent and develop random-matrix correlations.
We then track neutral-operator dynamics through operator-space entanglement and out-of-time-order correlations, and compare the resulting response scales with the spectral crossover.

Local charge memory and additional local probes are used to identify which observables distinguish the extent of the mismatch and which fail as general indicators of spectral chaos.
The perturbative interpretation and expanded architecture tests are presented separately in Secs.~\ref{Sec4_perturbative} and~\ref{Sec3p6_architecture}.

\subsection{Short-range level repulsion and persistent spectral rigidity}
\label{Sec3p1_spectral}
\begin{figure*}[htbp!]
\centering
\includegraphics[width=\linewidth]{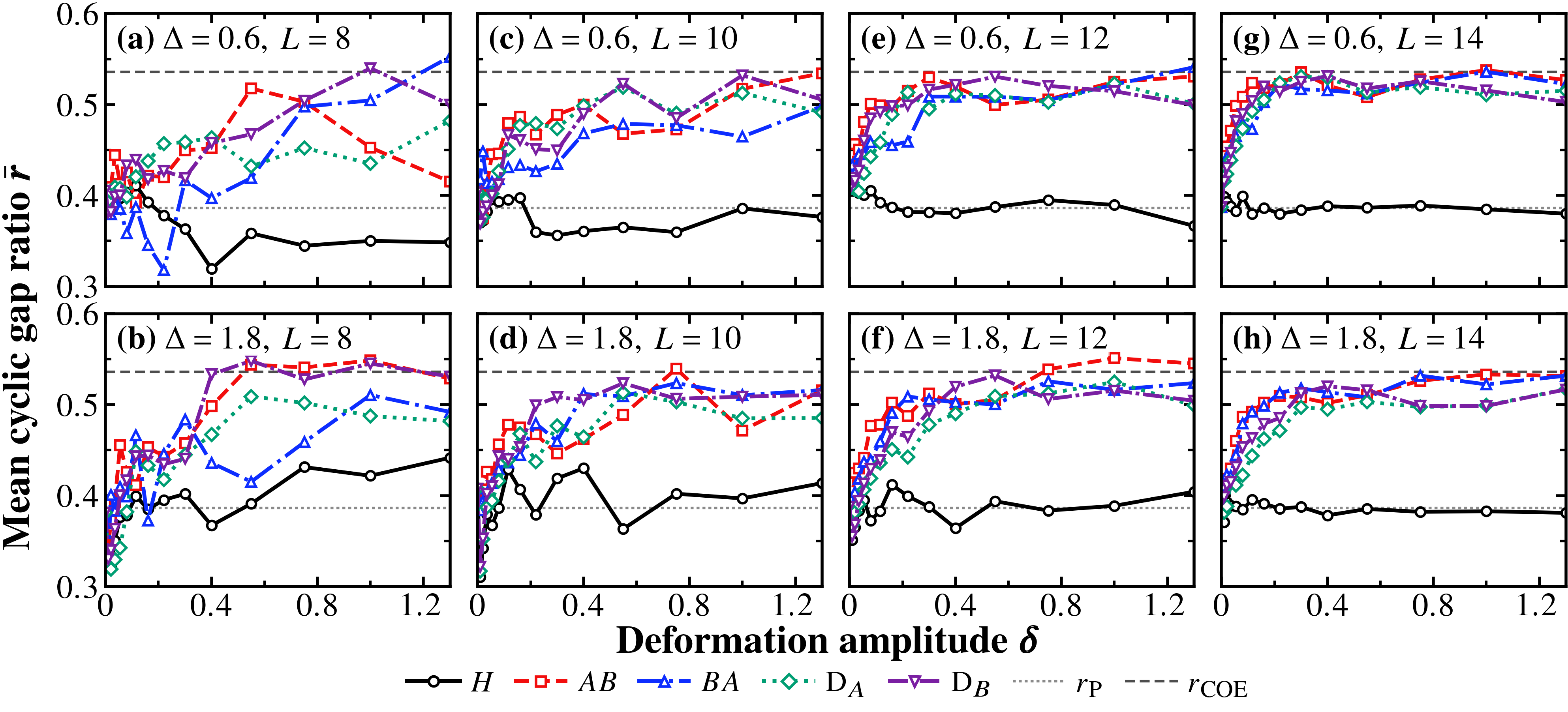}
\caption{Symmetry-resolved cyclic gap-ratio flow across contrast architectures and system sizes.
The mean cyclic adjacent-gap ratio $\bar{r}$ is shown as a function of $\delta$ for the XXZ parents $\Delta=0.6$ (top row) and $\Delta=1.8$ (bottom row), with $L=8,10,12,14$ from left to right.
The homogeneous reference $H$, full-support circuits $AB$ and $BA$, and single-defect circuits $D_A$ and $D_B$ are defined in Fig.~\ref{Fig_1}.
At $\delta=0$, the exact spin-flip and reflection blocks are resolved, whereas the full half-filled sector is used at positive $\delta$.
The dotted and dashed horizontal lines mark $r_{\text{P}}$ and $r_{\text{COE}}$, respectively.
The full-support flow toward COE shifts to smaller local amplitudes as $L$ increases, while the homogeneous reference remains predominantly Poisson-like.
Single defects also produce level repulsion, with less regular finite-size trajectories.}
\label{Fig_2}
\end{figure*}

We begin with the cyclic adjacent-gap ratio of the Floquet eigenphases, which probes spectral correlations between neighboring quasienergy levels without unfolding.
For each symmetry-resolved sequence containing $d$ eigenphases, we order them as $0\leq\theta_1\leq\cdots\leq\theta_d<2\pi$ and impose the cyclic conventions $\theta_{d+1}=\theta_1+2\pi$ and $s_{d+1}=s_1$.
The cyclic gaps are $s_n = \theta_{n+1}-\theta_n$, and the adjacent-gap ratios are defined by:
\begin{equation}
r_n = \frac{\min(s_n,s_{n+1})}{\max(s_n,s_{n+1})}\,, \quad
\bar{r} = \frac{1}{d}\sum_{n=1}^{d}r_n \,.
\end{equation}
We adopt standard Poisson and COE reference values $r_{\text{P}} = 2\ln(2)-1$, $r_{\text{COE}} \simeq 0.5359$.

At $\delta=0$, the ratios are evaluated separately within the exact $(F,R)$ blocks of the XXZ parent and then pooled.
For every positive $\delta$, the tested unitary symmetries are broken, and the full half-filled spectrum is used.
This symmetry treatment prevents independent spectral sequences from being mixed and artificially suppressing the gap ratio.

Fig.~\ref{Fig_2} shows that all five assignments coincide at $\delta=0$, as required by their common XXZ parent.
The homogeneous circuit remains predominantly Poisson-like throughout the scan and does not develop a systematic drift toward the COE benchmark with increasing $L$.
At $L=14$ and $\delta=1.3$, its gap ratios are $0.3800$ and $0.3810$ for $\Delta=0.6$ and $1.8$, respectively.
Because this reference already contains the uniform deformation generated by $W_{\text{avg}}$, its behavior shows that the common local deformation alone does not produce the level repulsion observed in the nonuniform circuits.

The homogeneous reference also clarifies the spatial origin of the response.
Repeating the same deformed gate preserves a uniform bond environment throughout the chain.
A single defect creates one localized scattering source that can hybridize extended many-body eigenstates.
The period-two circuits repeat the change of bond environment throughout the bulk and inject the contrast into a staggered spatial channel on every Floquet period.
The distinction is consequently between one dressed source region and a coherent array of sources.

The four nonuniform assignments develop a pronounced increase in $\bar{r}$ at positive $\delta$ for both XXZ parents.
The shift toward smaller deformation amplitudes as $L$ increases is clearest for the full-support circuits.
At $L=14$ and $\delta=1.3$, averaging the two full-support translations gives $\bar{r}=0.5246$ for $\Delta=0.6$ and $\bar{r}=0.5316$ for $\Delta=1.8$, both close to the COE value.
The single-defect circuits also develop substantial level repulsion, but their finite-size trajectories are more sensitive to the defect sign, placement, and deformation amplitude.

Fig.~\ref{Fig_2} shows family-dependent deformation scales, with no consistent pointwise ordering of the gap ratios.
As $L$ grows, a full-support pattern supplies additional dressed contrast insertions, whereas a defect remains associated with one spatial region.
Coherent cross terms can enhance or suppress the combined response of these sources.

The adjacent-gap ratio probes only the shortest spectral scale.
It can respond once neighboring eigenphases begin to repel, before the complete spacing distribution and correlations across several mean spacings have reorganized.
We next test whether the short-range flow in Fig.~\ref{Fig_2} develops into a persistent response of the full spectrum.

For the spacing-distribution test, the cyclic gaps in each symmetry-resolved sequence are rescaled to unit mean.
We compare them with the Poisson and COE reference cumulative distributions $F_{\text{P}}(s) = 1-\mathrm{e}^{-s}$ and $F_{\text{COE}}(s) = 1-\mathrm{e}^{-\pi s^2/4}$.
If $F_{\text{emp}}(s)$ denotes the empirical cumulative distribution, we define the Kolmogorov-Smirnov distances and their preference difference as $D_{\text{P}} = \sup_s|F_{\text{emp}}(s)-F_{\text{P}}(s)|$, $D_{\text{COE}} = \sup_s|F_{\text{emp}}(s)-F_{\text{COE}}(s)|$, and $P_{\text{KS}} = D_{\text{P}}-D_{\text{COE}}$.
A positive $P_{\text{KS}}$ means that the complete nearest-neighbor spacing distribution is closer to the COE reference than to the Poisson reference.

We complement this local spacing test with the circular number variance.
Let $N_X(x,\ell)$ count the globally rescaled eigenphases $x_n=d\theta_n/(2\pi)$ of circuit $X$ in the circular interval beginning at $x$ with length $\ell$.
For a sequence of $d$ levels, the uniform-origin number variance is:
\begin{equation}
\Sigma_X^2(\ell)=\frac{1}{d} \int_0^d [N_X(x,\ell)-\ell]^2\dd{x} \,.
\end{equation}
When exact symmetry blocks are present, the number variance is evaluated in each block and combined with the corresponding block dimensions.
At $\ell=5$, we quantify the rigidity gained by a nonuniform circuit relative to the homogeneous reference as $G_{\Sigma,X} = \Sigma_H^2(5)-\Sigma_X^2(5)$.
Thus $G_{\Sigma,X}>0$ indicates reduced level-count fluctuations and stronger spectral rigidity relative to $H$.

\begin{figure*}[htbp!]
\centering
\includegraphics[width=\linewidth]{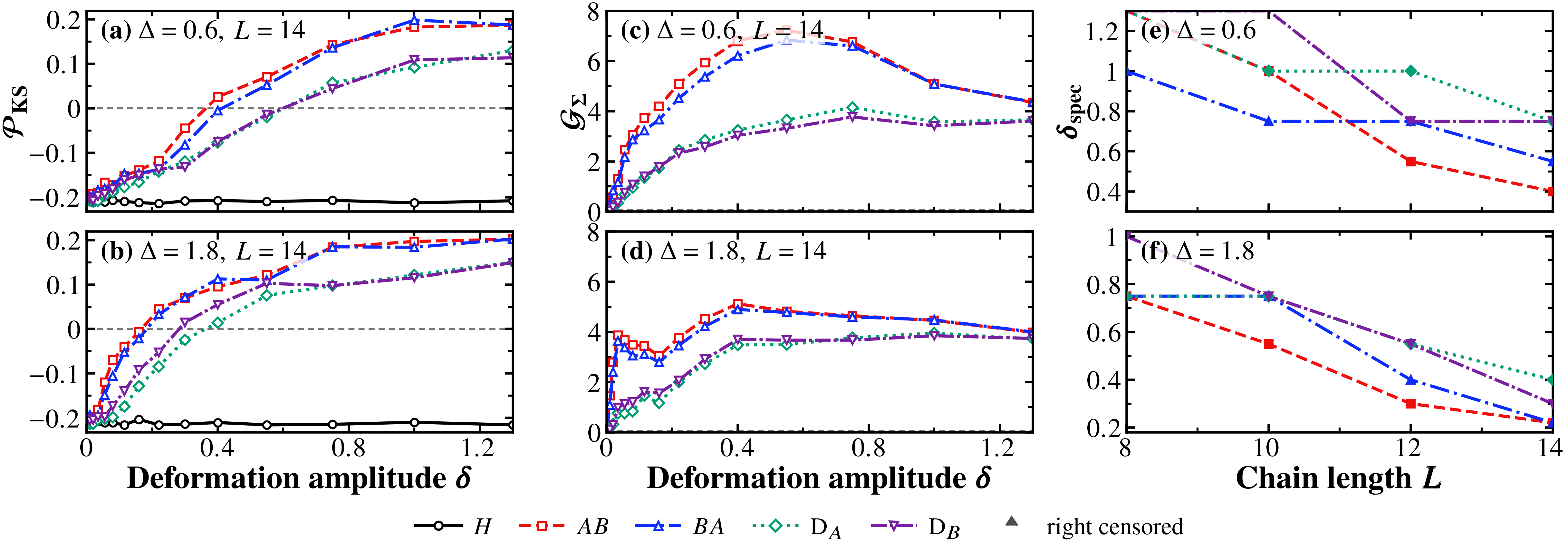}
\caption{Spacing distributions, spectral rigidity, and the joint spectral-response scale in finite-size circuits.
Panels (a) and (b) show the spacing-reference preference $P_{\text{KS}}$ at $L=14$ for $\Delta=0.6$ and $1.8$, respectively.
Panels (c) and (d) show the corresponding rigidity gain $G_{\Sigma}$ at $\ell=5$ relative to the shared homogeneous reference $H$.
Panels (e) and (f) show the persistent joint spectral-response scale $\delta_{\text{spec}}$ as a function of $L$.
The joint condition requires $\chi_{r,\text{env}}\geq0.5$, $P_{\text{KS}}>0$, and $G_{\Sigma}\geq0.05$ at two consecutive deformation points.
Upward carets mark right-censored cases for which no persistent joint response is found through $\delta=1.30$.
The positive spacing preference and rigidity gain confirm spectral reorganization beyond adjacent-level repulsion.
At $L=14$, both full-support circuits reach the joint criterion at smaller $\delta$ than either defect for each parent.}
\label{Fig_3}
\end{figure*}

To combine the gap-ratio, spacing-distribution, and rigidity responses, we first normalize the gap-ratio change of circuit $X$ by:
\begin{equation}
\chi_{r,X}(\delta) = \frac{\bar{r}_X(\delta)-\bar{r}_H(\delta)}{r_{\text{COE}}-\bar{r}_H(\delta)}\,,
\end{equation}
with $\chi_{r,\text{env},X}(\delta_i) = \max_{j\leq i}\chi_{r,X}(\delta_j)$.
The cumulative envelope prevents a later finite-size fluctuation from erasing an earlier gap-ratio response.
A deformation point $\delta_i$ satisfies the joint spectral condition when $\chi_{r,\text{env},X}(\delta_i) \geq 0.5$, $P_{\text{KS},X}(\delta_i) > 0$, and $G_{\Sigma,X}(\delta_i) \geq 0.05$.
We define $\delta_{\text{spec}}$ as the first grid point for which this condition also holds at the next sampled amplitude.
If no such consecutive pair occurs through $\delta=1.30$, the scale is reported as right-censored rather than assigned to the scan edge.

The spacing preference in Figs.~\ref{Fig_3}(a) and~\ref{Fig_3}(b) remains negative for the homogeneous circuit throughout the scan, with values close to $-0.21$ for both parents.
By contrast, all four nonuniform circuits cross to $P_{\text{KS}}>0$ and become COE-preferred at sufficiently large $\delta$.
Their positive rigidity gains in Figs.~\ref{Fig_3}(c) and~\ref{Fig_3}(d) show that this response extends beyond nearest-neighbor repulsion to suppressed level-count fluctuations over several mean spacings.
The rigidity condition is generally satisfied before the complete spacing distribution becomes COE-preferred, making the latter the limiting condition for the joint response at $L=14$.

The resulting persistent scales in Figs.~\ref{Fig_3}(e) and~\ref{Fig_3}(f) decrease or remain unchanged as the system size grows.
For $\Delta=0.6$ and $L=14$, the $AB$ and $BA$ scales are $0.40$ and $0.55$, whereas both single-defect scales are $0.75$.
For $\Delta=1.8$ at the same size, the two full-support scales are $0.22$, compared with $0.40$ for $D_A$ and $0.30$ for $D_B$.
At the largest tested size, both full-support circuits reach the joint spectral criterion before either defect for each parent.

The absolute scales differ between the two XXZ parents, demonstrating that source count acts together with the parent dynamics.
Because the contrast patterns are held fixed between the two parent anisotropies, this difference must enter through the parent-dependent quasienergy spacings and dressed contrast matrix elements.
The earlier crossover at $\Delta=1.8$ indicates a larger finite-size susceptibility to this particular contrast.

The comparison of Figs.~\ref{Fig_2} and~\ref{Fig_3} resolves a separation between short-range and broader spectral diagnostics on the finite-size grid.
Neighboring eigenphases can develop enhanced repulsion at amplitudes for which the full spacing distribution and longer-range rigidity have not yet responded persistently and consistently.
This separation is consistent with local hybridization of nearby quasienergy pairs preceding statistically typical mixing across several mean spacings.

The scale $\delta_{\text{spec}}$ is defined by the joint spectral criterion and its persistence across consecutive sampled amplitudes.
It characterizes the finite-size crossover on the sampled grid, rather than a thermodynamic transition point.
Having established the full-spectrum response, we next ask whether operator dynamics distinguish the same contrast architectures at lower deformation amplitudes.

\subsection{Neutral-operator response and comparison}
\label{Sec3p2_operator_response}
\begin{figure*}[htbp!]
\centering
\includegraphics[width=\linewidth]{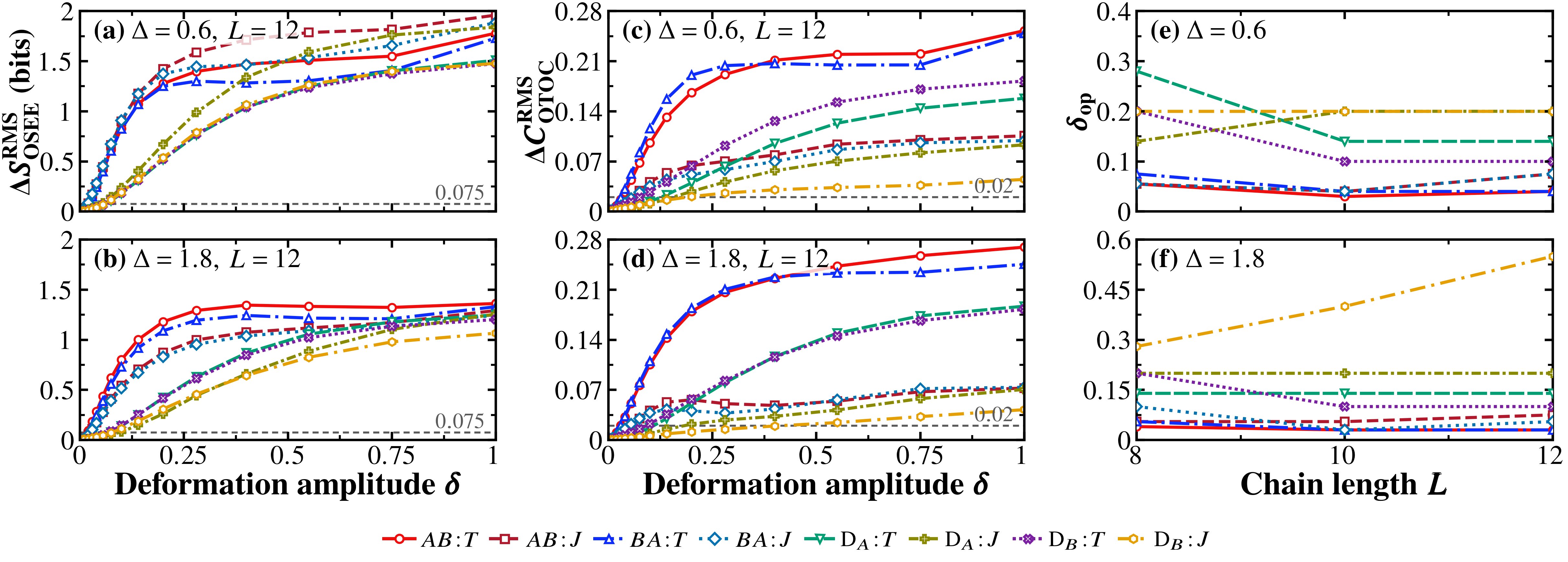}
\caption{Neutral-operator response across mismatch extent.
Panels (a) and (b) show the reference-relative RMS response of the charge-resolved half-chain OSEE at $L=12$ for $\Delta=0.6$ and $1.8$, respectively.
Panels (c) and (d) show the corresponding far-bond OTOC response.
The initial and far probes are both hopping operators $T$ or both current operators $J$.
The OSEE response is evaluated at $t=2,4,\ldots,48$, whereas the OTOC response is evaluated at $t=1,\ldots,48$.
Horizontal lines mark the thresholds $0.075$ bits and $0.020$.
Panels (e) and (f) show the persistent joint operator-response scale $\delta_{\text{op}}$ for $L=8,10,12$.
For both probes and parents, full support reaches the joint operator threshold at smaller $\delta$ than either defect at every shown size.
The RMS response measures the magnitude of dynamical changes, including either sign of the OTOC deviation.}
\label{Fig_4}
\end{figure*}
We probe the dynamics with the charge-neutral hopping and current operators $T_b$ and $J_b$, obtained by embedding the two-qubit operators $T$ and $J$ on bond $b$.
The initial operator $O_0$ is placed on the central bond $b_0=L/2$, which crosses the half-chain cut, and the corresponding far operator $O_{\text{far}}$ is placed on the right-edge bond $b_{\text{far}}=L-1$.
For each probe $O=T$ or $J$ and circuit $X$, the Heisenberg evolution is $O_{0,X}(t) = U_{F,X}^{-t}O_0U_{F,X}^t$.
Both probes preserve the total charge, so their evolution remains closed within the half-filled sector.

We first quantify operator spreading across the half-chain cut through the charge-resolved operator-space entanglement entropy~\cite{Prosen2007,Bertini2020a,Alba2025}.
The operator Schmidt decomposition is performed separately in sectors of the ket and bra charges of the left half-chain, after which all squared Schmidt singular values are normalized together.
If the resulting singular values are $s_\mu(t)$, we define $S_{\text{OSEE}}(t) = -\sum_\mu p_\mu(t)\log_2[p_\mu(t)]$, with:
\begin{equation}
p_\mu(t) = \frac{s_\mu^2(t)}{\sum_\nu s_\nu^2(t)} \,.
\end{equation}
For a nonuniform circuit $X$, let $T_e=\{2,4,\ldots,48\}$ denote the sampled even times.
Its root-mean-square OSEE response relative to the shared homogeneous reference $H$ is:
\begin{small}
\begin{equation}
\hspace{-0.5cm}
\Delta S_{\text{OSEE},X,\text{RMS}} = \sqrt{\frac{1}{24}\sum_{t\in T_e}[S_{\text{OSEE},X}(t)-S_{\text{OSEE},H}(t)]^2} \,.
\end{equation}
\end{small}

The complementary far-bond probe is the normalized squared commutator, a standard diagnostic of operator spreading and scrambling~\cite{Rampp2023,Dowling2023,Xu2024,Duh2026b}:
\begin{equation}
C^{\mathrm{comm}}_{O,X}(t) = \frac{d_N\norm{[O_{0,X}(t),O_{\text{far}}]}_F^2}{2\norm{O_0}_F^2\norm{O_{\text{far}}}_F^2} \,.
\end{equation}
The central and far bonds do not overlap, so $C^{\mathrm{comm}}_{O,X}(0)=0$.
The corresponding reference-relative OTOC response is:
\begin{small}
\begin{equation}
\Delta C_{\text{OTOC},X,\text{RMS}} = \sqrt{\frac{1}{48}\sum_{t=1}^{48}[C^{\mathrm{comm}}_{O,X}(t)-C^{\mathrm{comm}}_{O,H}(t)]^2} \,.
\end{equation}
\end{small}
Because this quantity squares the difference from the reference, it measures the magnitude of the change in far-bond operator propagation without fixing its direction.

We define the persistent operator-response scale $\delta_{\text{op}}$ as the first sampled amplitude at which $\Delta S_{\text{OSEE},X,\text{RMS}} \geq 0.075\,\text{bits}$, and $\Delta C_{\text{OTOC},X,\text{RMS}} \geq 0.020$.
Both inequalities must remain satisfied at the next deformation point.
No interpolation between sampled amplitudes is used.

Because all five primary circuits coincide with the undeformed parent at $\delta=0$, both RMS responses vanish there.
At positive amplitude, Figs.~\ref{Fig_4}(a)-\ref{Fig_4}(d) show that the finite-density circuits separate from the homogeneous reference much more rapidly than either single-defect circuit.
This separation appears in both the OSEE and far-bond OTOC channels and for both neutral probes.
It is strongest in the low-amplitude regime, before the full spectrum satisfies the persistent joint criterion of Fig.~\ref{Fig_3}.

\begin{figure*}[htbp!]
\centering
\includegraphics[width=\linewidth]{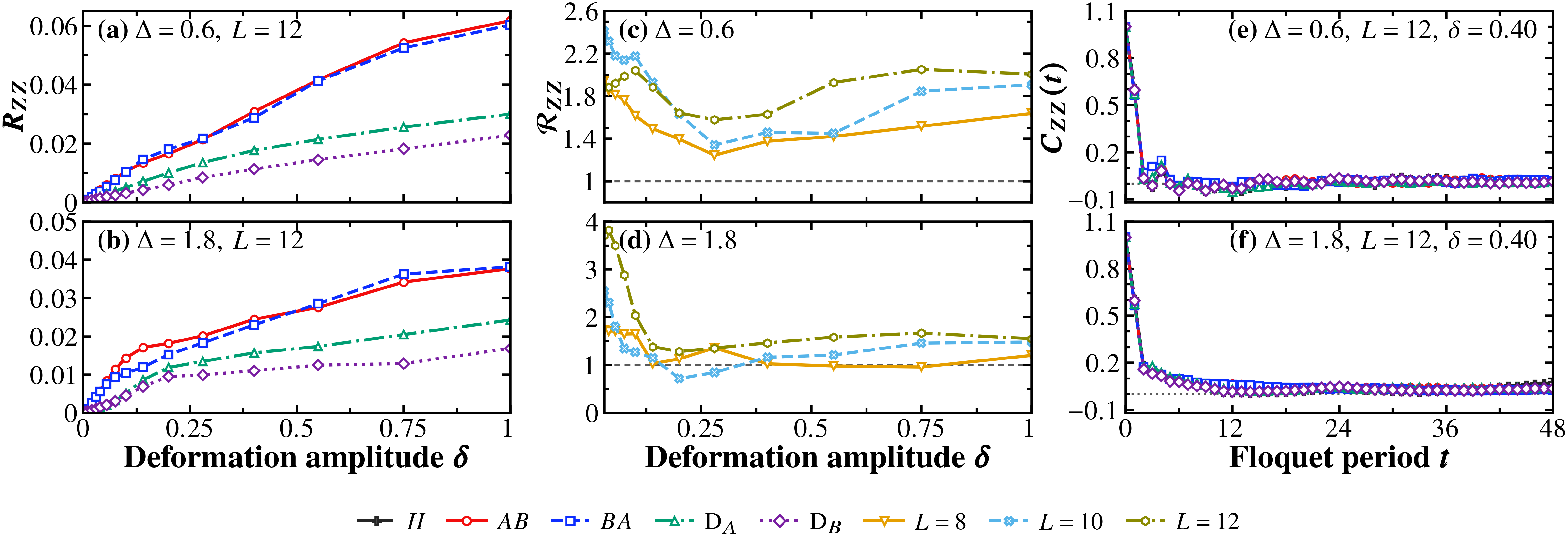}
\caption{Centered local-charge memory across mismatch extent.
Panels (a) and (b) show the reference-relative RMS response $R_{ZZ}$ at $L=12$ for $\Delta=0.6$ and $1.8$, respectively.
The measured site is $j_c=L/2$, the common endpoint of the two single-defect bonds.
Panels (c) and (d) show the conservative ratio $\mathcal{R}_{ZZ}$ for $L=8,10,12$ over the fixed analysis window $0.03\leq\delta\leq1$.
The dashed line marks $\mathcal{R}_{ZZ}=1$.
Panels (e) and (f) show representative autocorrelation traces at $L=12$ and $\delta=0.40$.
At $L=12$, $\mathcal{R}_{ZZ}>1$ at every sampled amplitude in the stated window for both parents, so local charge memory distinguishes the tested families.
A larger $R_{ZZ}$ measures a stronger change in the autocorrelation, which can include changes in oscillations and retained memory.}
\label{Fig_5}
\end{figure*}
This early sensitivity is consistent with contributions from additional contrast bonds as the operator spreads.
An initially local operator samples only gates inside its growing light cone.
In a defect circuit, the contrast enters this evolution through one fixed source region, even though reflections can generate repeated later encounters in a finite open chain.
In a finite-density circuit, the number of contrast bonds inside the operator support grows together with the support itself.
Repeated interactions with these contrast bonds can modify the operator's Schmidt structure and its commutators with distant probes during early propagation.
For the small open chains studied here, the full window through $t=48$ also contains boundary returns, saturation, and recurrences.

The operational scales in Figs.~\ref{Fig_4}(e) and~\ref{Fig_4}(f) make the family hierarchy quantitative.
All finite-density response scales lie between $0.03$ and $0.10$, whereas the single-defect response scales lie between $0.10$ and $0.55$.
For all 12 combinations of parent, system size, and probe, both finite-density circuits respond at lower amplitudes than either defect circuit.
The corresponding defect-to-finite-density scale ratios range from $2$ to $5$, with a median of $2.67$.
Thus the distributed contrast reorganizes neutral-operator dynamics at substantially lower local deformation amplitudes than a single contrast defect over the full tested operator-dynamics grid.

The OSEE difference is predominantly positive, indicating that the nonuniform circuits usually generate greater operator-space entanglement than the homogeneous reference.
The signed OTOC difference, however, is predominantly positive for the hopping probe and predominantly negative for the current probe.
The RMS OTOC response establishes a probe-dependent modification of operator propagation, not a universal enhancement of scrambling.

The operators $T$ and $J$ are two Hermitian quadratures of the same local charge-transfer channel.
Since the contrast generators contain both quadratures, coherent evolution can redistribute amplitude and phase between hopping-like and current-like components.

At the original thresholds, operator dynamics generally respond at lower deformation amplitudes than the spectral diagnostics.
Among the 40 comparisons for which $\delta_{\text{spec}}$ is resolved, 39 satisfy $\delta_{\text{op}}<\delta_{\text{spec}}$, while the remaining comparison places the two scales at the same grid point.
The spectral scale is right censored in the remaining cases even though the operator scale is already resolved.

This ordering indicates that a propagating operator can become sensitive to the spatial extent of the contrast before the complete quasienergy spectrum acquires a persistent COE-type response.
Across the expanded sample, the operator onset remains below the joint spectral onset at $t_{\text{max}}=16,24,48$ for all tested threshold variations.
The ordering can change at longer times or with a different response criterion.
Even a homogeneous integrable deformation produces appreciable operator changes, so an early response need not imply chaotic spectral correlations (Appendix~\ref{SecS5_architecture}).
We next test whether the same mismatch-extent information remains visible in the memory of a single local charge.

\subsection{Mismatch extent test using local charge memory}
\label{Sec3p3_local_charge}
We measure the local charge on the central site $j_c=L/2$, which is the common endpoint of the two defect bonds; local measurements can provide sensitive probes of many-body chaos~\cite{VallejoFabila2025a}.
This choice keeps the probe adjacent to both tested defects and avoids a trivial difference in probe-defect distance.
Let $P_N$ project onto the half-filled sector and let $\Tr_N$ denote the trace within that sector.
We remove the fixed-sector background by defining:
\begin{equation}
\tilde{Z}_{j_c} = P_NZ_{j_c}P_N-\frac{\Tr_N(P_NZ_{j_c}P_N)}{d_N}\mathbb{1}_N \,.
\end{equation}
The subtracted sector mean vanishes for the half-filled chains studied here.

For circuit $X$, the normalized infinite-temperature autocorrelation is:
\begin{equation}
C_{ZZ,X}(t) = \frac{\Tr_N[\tilde{Z}_{j_c,X}(t)\tilde{Z}_{j_c}]}{\Tr_N[\tilde{Z}_{j_c}^2]} \,,
\end{equation}
where $\tilde{Z}_{j_c,X}(t) = U_{F,X}^{-t}\tilde{Z}_{j_c}U_{F,X}^t$.
This normalization gives $C_{ZZ,X}(0)=1$.
The reference-relative response over the first 48 Floquet periods is:
\begin{equation}
R_{ZZ,X} = \sqrt{\frac{1}{48}\sum_{t=1}^{48}[C_{ZZ,X}(t)-C_{ZZ,H}(t)]^2} \,.
\end{equation}
As in the operator diagnostics, the homogeneous circuit is evaluated at the same $\Delta$, $L$, and $\delta$ as circuit $X$.

We compare the mismatch families using the following conservative response ratio:
\begin{equation}
\mathcal{R}_{ZZ} = \frac{\min(R_{ZZ,AB},R_{ZZ,BA})}{\max(R_{ZZ,D_A},R_{ZZ,D_B})} \,.
\end{equation}
The numerator is the weaker of the two finite-density responses, while the denominator is the stronger of the two single-defect responses.
Consequently, $\mathcal{R}_{ZZ}>1$ means that the finite-density family produces the larger local-memory response even under the least favorable tested placement comparison.

Figs~\ref{Fig_5}(a) and~\ref{Fig_5}(b) show a clear separation at the largest dynamical size.
For $L=12$, both finite-density responses exceed both single-defect responses at every sampled amplitude in the fixed analysis window for each XXZ parent.
The corresponding ratios in Figs.~\ref{Fig_5}(c) and~\ref{Fig_5}(d) remain above unity for all 24 combinations of parent and amplitude at this size.
Thus a measurement at one central site retains information about whether the contrast is localized on one bond or distributed across the circuit.

Both bonds adjacent to the central site are mismatched in the finite-density circuits, whereas only one adjacent bond is mismatched in either defect circuit.
The early-time response can distinguish the immediate local gate environments without sampling the global circuit.
At later times, $C_{ZZ}(t)$ also contains propagation away from the central site, boundary returns, and coherent paths that encounter more distant sources.
The time-integrated RMS response combines local and nonlocal contributions, so family discrimination alone does not establish a local readout of global mismatch extent.

The smaller sizes reveal the boundary of this local discriminator.
For $\Delta=0.6$, all 36 combinations of system size and deformation amplitude satisfy $\mathcal{R}_{ZZ}>1$.
For $\Delta=1.8$, the criterion holds in 32 of 36 combinations, with exceptions at $(L,\delta)=(8,0.55)$, $(8,0.75)$, $(10,0.20)$, and $(10,0.28)$.
Overall, 68 of the 72 conservative comparisons favor the finite-density response.
Whether the local observable distinguishes the architectures depends on the parent, size, amplitude, and the tested signs and positions of the contrast sources.
Interference between propagation paths, boundary returns, and parent-dependent phases may contribute to these reversals.

The representative traces in Figs.~\ref{Fig_5}(e) and~\ref{Fig_5}(f) clarify what this response measures.
The five circuits share the initial value $C_{ZZ}(0)=1$ but differ in their early decay, overshoot, oscillation phase, and residual late-time memory.
Since the difference is squared, a larger $R_{ZZ}$ can result from either less or more retained charge memory, as well as from a reshaped oscillatory trajectory.
It measures the local distinguishability of the circuit families, not a decay rate or an independent degree of quantum chaos.

We next determine how strongly this conclusion depends on the choice of local observable and whether a local correlation hole supplies any additional chaos diagnostic.

\subsection{Probe dependence and the limits of local chaos diagnostics}
\label{Sec3p4_probe_dependence}
We extend the local comparison to the central-bond hopping and current operators and to the two-site one-particle projector $\Pi_1 = \ketbra{01}{01}+\ketbra{10}{10}$.
The projector is embedded on the central bond, projected into the half-filled sector, and centered by the same sector-trace subtraction used for the local charge.
We write $O_X(t)=U_{F,X}^{-t}OU_{F,X}^t$ for the evolution of a local probe under circuit $X$.
For each centered or traceless local probe $O$, we define:
\begin{subequations}
\begin{equation}
C_{O,X}(t) = \frac{\Tr_N[O_X(t)O]}{\Tr_N(O^2)} \,.
\end{equation}
\begin{equation}
R_{O,X} = \sqrt{\frac{1}{48}\sum_{t=1}^{48}[C_{O,X}(t)-C_{O,H}(t)]^2} \,.
\end{equation}
\end{subequations}

\begin{figure*}[htbp!]
\centering
\includegraphics[width=\linewidth]{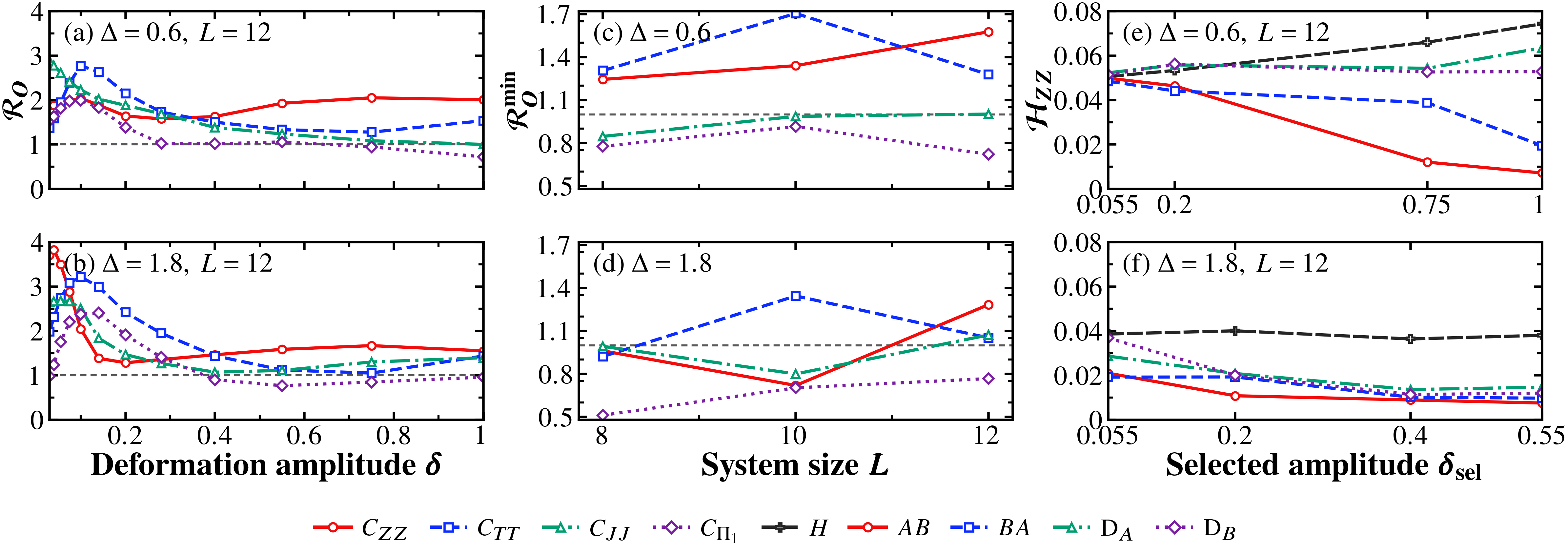}
\caption{Probe dependence of local mismatch discrimination and the non-diagnostic character of correlation holes.
Panels (a) and (b) show the conservative family ratio $\mathcal{R}_O$ at $L=12$ for the centered charge, hopping, current, and centered one-particle projector probes.
Panels (c) and (d) show the minimum ratio $\mathcal{R}_{O,\text{min}}$ over the sampled deformation amplitudes as a function of $L$.
The horizontal line at unity separates contexts in which the weaker full-support response exceeds the stronger tested defect response.
Panels (e) and (f) show the normalized positive correlation-hole depth $\mathcal{H}_{ZZ}$ at $L=12$ for amplitudes selected from the operator and spectral response regimes.
The plateaus are evaluated exactly, and the finite-time minima are taken over $t=1,\ldots,240$.
At $L=12$, the minimum ratios exceed unity for the charge, hopping, and current probes for both parents, but fall below unity for the projector.
Positive correlation holes also occur in the integrable homogeneous reference, so their presence alone does not diagnose chaos here.}
\label{Fig_6}
\end{figure*}
The four probe channels are denoted by $C_{ZZ}$, $C_{TT}$, $C_{JJ}$, and $C_{\Pi_1}$.
Let $\mathcal{G}_\delta$ denote the sampled deformation grid.
We define the conservative family ratios and their minima over the deformation grid as:
\begin{subequations}
\begin{equation}
\mathcal{R}_O = \frac{\min(R_{O,AB},R_{O,BA})}{\max(R_{O,D_A},R_{O,D_B})} \,,\\
\end{equation}
\begin{equation}
\mathcal{R}_{O,\text{min}}(L,\Delta) = \min_{\delta\in\mathcal{G}_\delta\cap[0.03,1]}\mathcal{R}_O(L,\Delta,\delta) \,.
\end{equation}
\end{subequations}
We separately test whether a dip below the infinite-time value supplies information beyond the reference-relative response.

For a finite Floquet spectrum, the exact infinite-time plateau is:
\begin{equation}
C_{ZZ,X,\infty} = \lim_{n_{\text{av}}\to\infty}\frac{1}{n_{\text{av}}}\sum_{t=0}^{n_{\text{av}}-1}C_{ZZ,X}(t) \,.
\end{equation}
It is evaluated from the quasienergy eigenspaces, including exact degeneracies, rather than estimated from a finite-time tail.
Using traces through $t=240$, we define the normalized positive correlation-hole depth as:
\begin{subequations}
\begin{equation}
h_{ZZ,X,+} = \max(C_{ZZ,X,\infty}-\min_{1\leq t\leq240}C_{ZZ,X}(t),0) \,,
\end{equation}
\begin{equation}
\mathcal{H}_{ZZ,X} = \frac{h_{ZZ,X,+}}{|1-C_{ZZ,X,\infty}|} \,.
\end{equation}
\end{subequations}
A positive $\mathcal{H}_{ZZ,X}$ records that the finite-time trace falls below its exact plateau.

The reason this dip is not a purely spectral observable is explicit in the Floquet eigenbasis.
For a Hermitian local probe with matrix elements $O_{\alpha\beta,X}=\mel{\alpha}{O}{\beta}$, its autocorrelation has the representation:
\begin{equation}
C_{O,X}(t) = \frac{1}{\Tr_N(O^2)}\sum_{\alpha,\beta}\mathrm{e}^{\mathrm{i}(\theta_{\alpha,X}-\theta_{\beta,X})t}|O_{\alpha\beta,X}|^2 \,.
\end{equation}
The time dependence combines quasienergy differences with probe-dependent matrix-element weights.
Consequently, a local correlation hole depends on eigenstate structure and conservation-induced selection rules in addition to spectral correlations.

The local response depends on the probe, as shown in Figs.~\ref{Fig_6}(a) and~\ref{Fig_6}(b).
At $L=12$, the minimum ratios for $C_{ZZ}$, $C_{TT}$, and $C_{JJ}$ remain above unity for both parents.
The minimum for $C_{\Pi_1}$ falls below unity in each case.
Across both parents and all three sizes, the weaker full-support response exceeds the stronger tested defect response in 71 of 72 comparisons for $C_{TT}$.
The corresponding counts are 68 for $C_{ZZ}$, 67 for $C_{JJ}$, and 48 for $C_{\Pi_1}$.
The hopping autocorrelation is the most stable additional local discriminator on the tested grid, while the projector lacks comparable robustness.

This ordering may reflect the overlap between the probes and deformation generators.
The deformation generators contain hopping-like and current-like coherent components, so $T$ and $J$ directly sample channels modified by the mismatch.
The projector $\Pi_1$ detects whether the central bond occupies its one-particle sector but does not resolve the coherent orientation within that sector.
Locality and charge conservation alone are insufficient to guarantee sensitivity to mismatch architecture.
The finite-size failures of $C_{JJ}$ and $C_{\Pi_1}$ further rule out a universal ordering of local probes.

All 40 traces in Figs.~\ref{Fig_6}(e) and~\ref{Fig_6}(f) have a positive correlation hole across the selected circuits, parents, and amplitudes, including every homogeneous reference.
Moreover, the homogeneous circuit has the deepest hole among the five circuits in six of the eight combinations of parent and amplitude.
This behavior is opposite to any simple identification of a deeper local hole with the stronger COE-type spectral response found for the nonuniform circuits.

A local correlation hole depends on spectral rigidity, the finite Hilbert space, charge conservation, boundary reflections, quasienergy degeneracies, and recurrences.
In these circuits, a positive local correlation hole is insufficient to diagnose spectral chaos.
Together, Figs.~\ref{Fig_5} and~\ref{Fig_6} show that local observables can distinguish the tested contrast architectures while remaining nonuniversal indicators of global spectral behavior.

\section{\texorpdfstring{Perturbative interpretation of the architecture-dependent response}{Perturbative interpretation of the architecture-dependent response}}
\label{Sec4_perturbative}
The preceding results show that full-support and single-defect circuits respond at different local deformation amplitudes.
We now use a bond-resolved perturbative construction to separate the accumulated contributions of individual contrast sources from coherent cross terms between different bonds.

\subsection{\texorpdfstring{Bond-resolved contrast susceptibility}{Bond-resolved contrast susceptibility}}
\label{Sec2p6_tangent}
For a primary architecture $\bm{s}$, an auxiliary parameter $\lambda$ connects the reference $H$ to the physical circuit through $u_b(\delta,\lambda) = \mathrm{e}^{-\mathrm{i}\delta(W_{\text{avg}}+\lambda s_bW_c)}u_0$ with $0\leq\lambda\leq1$.
The point $\lambda=0$ gives $H$, while $\lambda=1$ gives the chosen contrast architecture.
Varying $\lambda$ at fixed $\delta$ changes only the contrast, keeping the homogeneous reference fixed; the physical $\delta$ scan changes both the common and contrast deformations.
The auxiliary expansion isolates the response to the spatial mismatch about this reference.
Differentiation with respect to $\lambda$ at fixed $\delta$ gives the local contrast tangent:
\begin{equation}
W_c(\delta) = \int_0^1\mathrm{e}^{-\mathrm{i}x\delta W_{\text{avg}}}W_c\mathrm{e}^{\mathrm{i}x\delta W_{\text{avg}}}\dd{x} \,.
\end{equation}

We order the $M$ gate slots within one Floquet period by $\ell=1,\ldots,M$ and let $b_\ell$ be the bond at slot $\ell$.
We denote the embedded reference gate at that slot by $h_\ell(\delta)$.
The product of gates acting after slot $\ell$ and the dressed one-period contrast tangent are:
\begin{subequations}
\begin{align}
A_\ell(\delta) &= h_M(\delta)h_{M-1}(\delta)\cdots h_{\ell+1}(\delta) \,,\\
K_{\bm{s}}(\delta) &= \sum_{\ell=1}^Ms_{b_\ell}A_\ell(\delta)W_{c,b_\ell}(\delta)A_\ell^\dagger(\delta) \,.
\end{align}
\end{subequations}
Here $W_{c,b_\ell}$ denotes the local tangent embedded on bond $b_\ell$.
To first order in $\lambda$, the Floquet operator becomes:
\begin{equation}
U_{\bm{s}}(\delta,\lambda) = [\mathbb{1}-\mathrm{i}\lambda\delta K_{\bm{s}}(\delta)+\mathcal{O}(\lambda^2\delta^2)]U_H(\delta) \,.
\end{equation}

For a local probe $O$, we define $O_H(t)=U_H^{-t}OU_H^t$ and $K_{\bm{s}}(m;\delta)=U_H^{-m}K_{\bm{s}}(\delta)U_H^m$.
Its first-order response $O_{\bm{s}}(t,\lambda)-O_H(t)$ is:
\begin{equation}
\mathrm{i}\lambda\delta\sum_{m=1}^t[K_{\bm{s}}(m;\delta),O_H(t)]+\mathcal{O}(\lambda^2\delta^2) \,.
\end{equation}
Each commutator receives contributions only from dressed sources whose light cones overlap the evolving probe.
A single defect remains tied to one source region, whereas a full-support architecture can contribute through progressively more bonds as the probe spreads.
This repeated exposure is consistent with the smaller $\delta_{\text{op}}$ of the full-support circuits during early propagation.

The same tangent defines a quasienergy-mixing susceptibility.
After removing the fixed-sector identity component, we write:
\begin{equation}
K_{\bm{s},\circ} = P_NK_{\bm{s}}P_N-\frac{\Tr_N(P_NK_{\bm{s}}P_N)}{d_N}\mathbb{1}_N \,.
\end{equation}
Let $K_{\ell,\circ}$ be the centered tangent obtained by retaining only slot $\ell$.
In the eigenbasis of $U_H$, the full and self-only off-diagonal variances are:
\begin{subequations}
\begin{align}
v_{\bm{s},\text{full}} &= \frac{1}{d_N(d_N-1)}\sum_{\alpha\neq\beta}|(K_{\bm{s},\circ})_{\alpha\beta}|^2 \,,\\
v_{\bm{s},\text{self}} &= \frac{1}{d_N(d_N-1)}\sum_{\alpha\neq\beta}\sum_{\ell=1}^Ms_{b_\ell}^2|(K_{\ell,\circ})_{\alpha\beta}|^2 \,.
\end{align}
\end{subequations}
The self-only variance provides an additive-source baseline.
It retains the individual off-diagonal weight of each dressed bond but discards interference between different bonds.
The full variance restores these cross terms, so comparing the two quantities measures their contribution to this spectral susceptibility.

We denote their difference by $v_{\bm{s},\text{cross}}=v_{\bm{s},\text{full}}-v_{\bm{s},\text{self}}$.
Comparing the root-mean-square off-diagonal tangent element with the mean quasienergy spacing $2\pi/d_N$ gives:
\begin{equation}
\delta_{\text{mix},\bm{s},q} = \frac{2\pi}{d_N\sqrt{v_{\bm{s},q}}}, \quad q\in\{\text{self},\text{full}\} \,.
\end{equation}
At fixed sector dimension, a larger off-diagonal variance gives a smaller $\delta_{\text{mix}}$, indicating greater susceptibility to quasienergy mixing within this estimate.

The subscript $\text{FD}$ denotes the arithmetic mean over $AB$ and $BA$, while $\text{D}$ denotes the mean over $D_A$ and $D_B$.
If individual bulk-source weights are comparable, the $M$ self terms of a full-support architecture predict:
\begin{equation}
\frac{\delta_{\text{mix},\text{FD},\text{self}}}{\delta_{\text{mix},\text{D},\text{self}}} \simeq \frac{1}{\sqrt{M}} \,.
\end{equation}
This relation concerns the self-only tangent susceptibility and is not a scaling law for the nonlinear spectral or operator response.

To determine whether the all-pair average hides a different response among nearby levels, we also evaluate the same full and self-only weights for cyclically adjacent eigenphases, for phase pairs whose circular phase distance does not exceed $0.1\pi$, and with inverse-spacing weights regularized by the mean quasienergy spacing (Appendix~\ref{SecS2_mixing}).

\subsection{\texorpdfstring{Source-count scaling and coherent corrections}{Source-count scaling and coherent corrections}}
\label{Sec3p5_physical_origin}
We evaluate the bond-resolved susceptibility defined in Sec.~\ref{Sec2p6_tangent} at the homogeneous reference amplitudes $\delta_{\text{ref}}=0.02$, $0.04$, and $0.08$.
These values specify the linearization point and are not fitted to the measured crossover scales.
Because $\delta_{\text{mix}}$ uses all off-diagonal matrix elements and the mean quasienergy spacing, it measures relative susceptibility rather than the absolute onset of level repulsion.
The rapidly decreasing many-body quasienergy spacing can increase the susceptibility of both architectures with size, even as the single-defect density vanishes.

\begin{figure}[!t]
\centering
\includegraphics[width=\columnwidth]{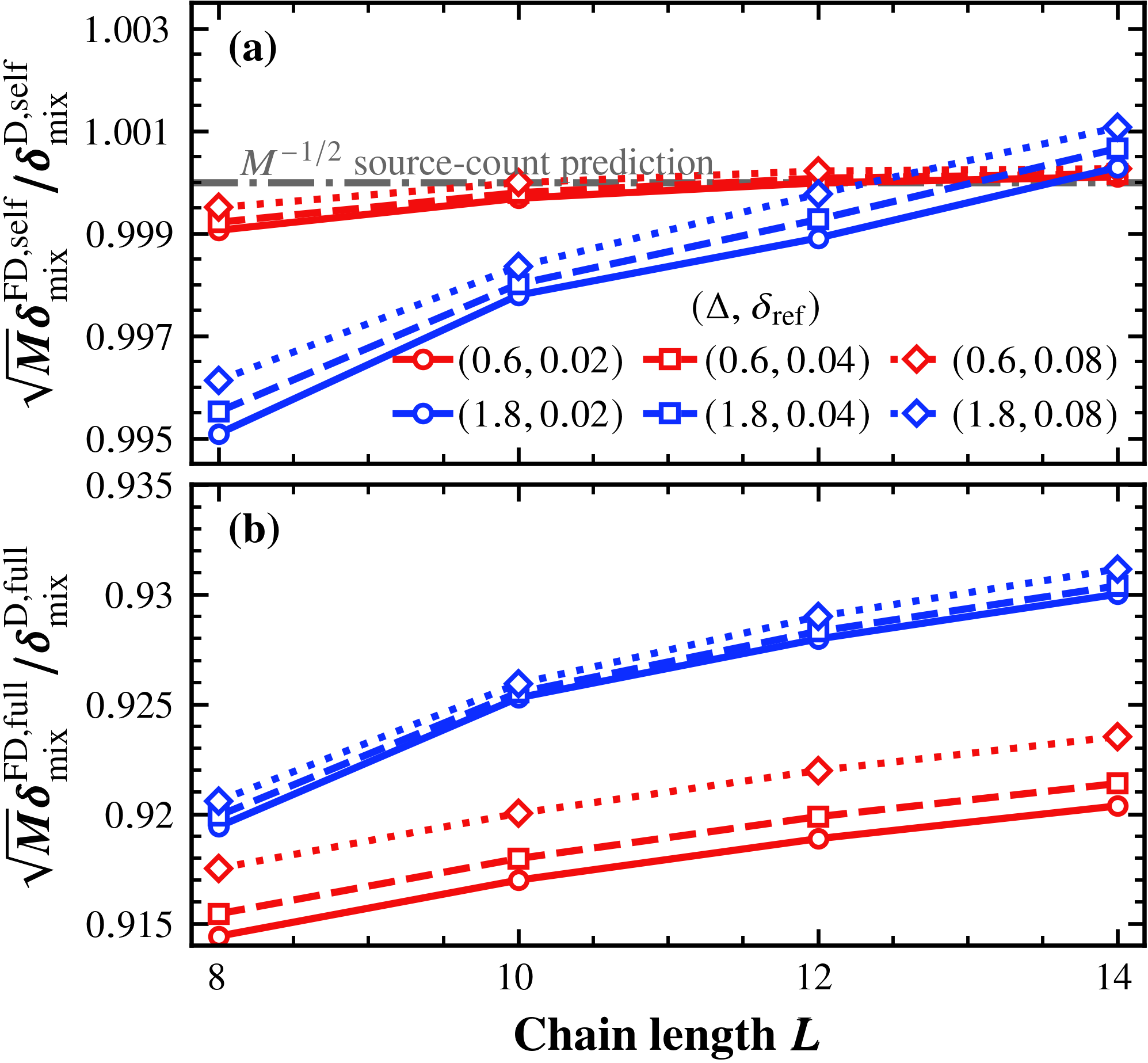}
\caption{Source-count scaling of the perturbative quasienergy-mixing estimate.
Here $M=L-1$ is the number of mismatch sources in a finite-density circuit.
The subscripts $\text{FD}$ and $\text{D}$ denote arithmetic means over $AB$ and $BA$, and over $D_A$ and $D_B$, respectively.
Panel (a) shows the self-only mixing-scale ratio after rescaling by $\sqrt{M}$.
Each legend entry gives the pair $(\Delta,\delta_{\text{ref}})$, and the gray dash-dotted line marks the $M^{-1/2}$ source-count prediction.
Panel (b) shows the corresponding ratio after restoring all coherent bond-bond cross terms.
The self-only data follow $M^{-1/2}$, while coherent cross terms lower this all-pair mixing estimate by about $7\%$-$9\%$.
This scaling describes the perturbative susceptibility and does not establish a scaling law for the measured nonlinear crossover.}
\label{Fig_7}
\end{figure}

Fig.~\ref{Fig_7}(a) shows that the self-only data collapse onto the $M^{-1/2}$ prediction for both XXZ parents and all three reference amplitudes.
The rescaled ratio lies between $0.9951$ and $1.0011$ over the complete grid, so deviations from the source-count law remain below $0.5\%$.
The collapse indicates comparable self-only weights for the bulk sources and selected central defects.
Restoring coherent cross terms shifts the rescaled ratio to $0.9144$-$0.9312$, as shown in Fig.~\ref{Fig_7}(b).
The finite-density cross fraction $v_{\bm{s},\text{cross}}/v_{\bm{s},\text{full}}$ lies between $0.135$ and $0.162$, so interference is constructive in this off-diagonal tangent variance and lowers its mixing-scale estimate by approximately $7\%$-$9\%$ relative to the self-only baseline.
The self-only and full estimates give identical Spearman orderings relative to the measured $\delta_{\text{spec}}$ at all three reference amplitudes, with coefficients $0.880$ for $\Delta=0.6$ and $0.917$ for $\Delta=1.8$.
Within this spectral tangent metric, additive source weight dominates the architecture separation, with a smaller correction from coherent bond-bond interference.

The all-pair average does not determine how the closest quasienergy levels are coupled.
The near-level analysis in Appendix~\ref{SecS4_null} repeats the full and self-only decomposition for adjacent eigenphases, for phase pairs whose circular phase distance does not exceed $0.1\pi$, and for a spacing-weighted susceptibility.
The adjacent-pair weight is smaller than the all-pair weight for every size at $\Delta=0.6$, but larger by a factor of $2.02$-$4.17$ at $\Delta=1.8$.
The adjacent cross fraction is positive, whereas the fixed-window and spacing-weighted cross fractions can have either sign.

Moreover, the Spearman correlation between adjacent gaps and coupling strengths remains between $-0.005$ and $0.106$.
Interference redistributes pair-resolved mixing in a parent- and size-dependent way without producing a universal enhancement of the nearest quasienergy pairs.
In the separate fixed-budget analysis at a common background, typical adjacent-level channels show strong cancellation that can be obscured in averages by a minority of enhanced channels (Appendix~\ref{SecS5_architecture}).

\section{\texorpdfstring{Architecture robustness and diagnostic dependence}{Architecture robustness and diagnostic dependence}}
\label{Sec3p6_architecture}
We now test whether the endpoint hierarchy survives beyond the original generator pair and whether placement matters after source count and lower spatial moments are held fixed.
The expanded and strictly matched architecture families are defined in Sec.~\ref{Sec2p3_architectures}.

The spectral scan reaches $L=14$, and the operator scan reaches $L=12$ with several time windows and thresholds.
For the broader spectral comparison, $\delta_{r,50}$ is the interpolated amplitude at which the running envelope of the normalized gap-ratio response first reaches $0.5$.
This scale remains observable for most generator-architecture cases, unlike the stricter persistent joint criterion, and no right-censored crossing is assigned to the scan boundary.
All $560$ cases tested at $L=8$ satisfy the same antiunitary symmetry relation as the primary family, so the comparisons remain within one circular-ensemble symmetry class.

\begin{figure*}[htbp!]
\centering
\includegraphics[width=\linewidth]{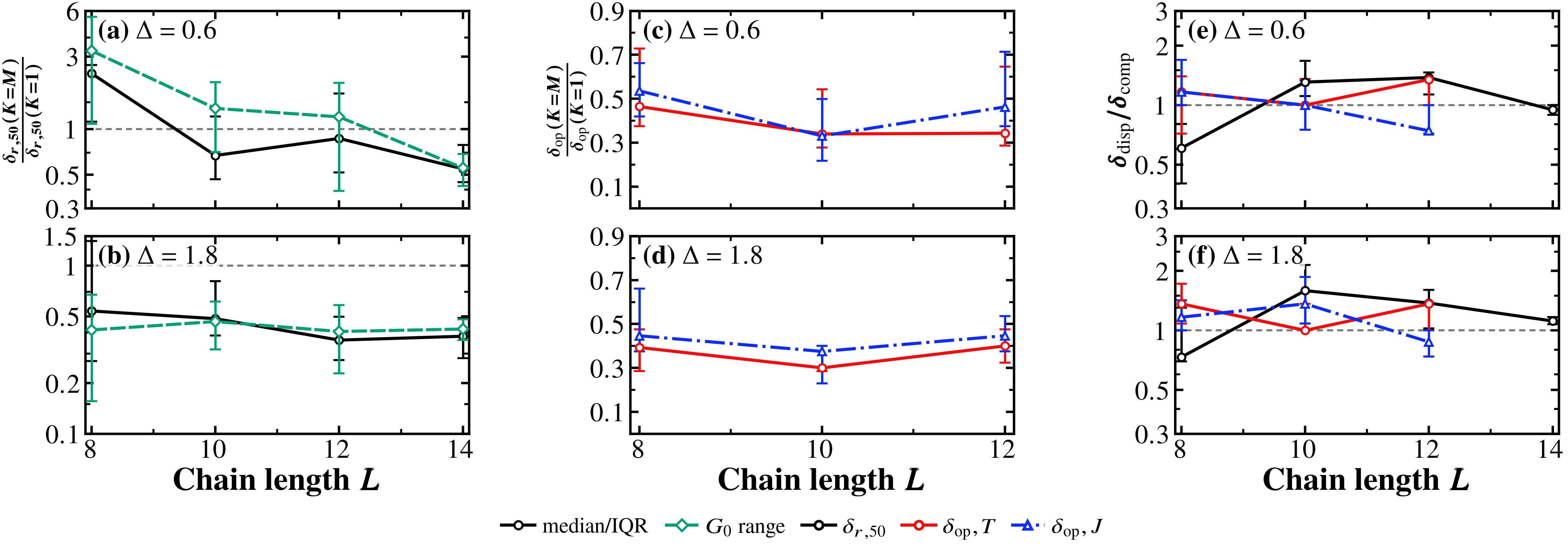}
\caption{Architecture robustness and diagnostic dependence of the response hierarchy.
Each architecture response is measured relative to the sign-matched homogeneous control with the same bond-averaged generator.
Panels (a) and (b) show the ratio $\delta_{r,50}(K=M)/\delta_{r,50}(K=1)$ for $\Delta=0.6$ and $1.8$, respectively.
Each ratio pairs architectures at the same generator, parent, size and starting sign; values below unity indicate an earlier full-support response.
Black circles give the median over the five fixed generator pairs and both starting signs, with interquartile error bars.
These bars span the $25$th-$75$th percentiles across generator-sign choices, quantifying sample variation.
Green diamonds give the median for the original pair $G_0$, with the range over its two starting signs.
Panels (c) and (d) show the corresponding ratio of the persistent absolute operator-response scales at $t_{\text{max}}=48$ for the hopping probe $T$ and current probe $J$.
Panels (e) and (f) compare the layer- and moment-matched placement pairs through the dispersed-to-compact response-scale ratio.
Black circles show $\delta_{r,50}$, while red circles and blue triangles show the persistent absolute scales for $T$ and $J$ at $t_{\text{max}}=48$.
Only comparisons in which both scales are observed enter the ratios.
The endpoint medians favor full support at fixed local amplitude, for operator dynamics at every shown size and for spectral response at $L=14$.
The matched placement ratios cross unity, so neither compact nor dispersed patterns respond earlier throughout the tested cases.}
\label{Fig_8}
\end{figure*}

The operator endpoint hierarchy is stable across the five fixed generator pairs.
The $120$ comparisons cover generator pairs, parents, system sizes, starting signs, and probes.
The $K=M$ response scale is lower than the $K=1$ scale in $109$ cases, equal in $7$, and higher in $4$.
The median ratios in Figs.~\ref{Fig_8}(c) and~\ref{Fig_8}(d) remain well below unity for both probes, parents, and all three sizes.

The spectral endpoint comparison is less uniform at the smaller sizes.
Using observed onsets and censoring bounds, the full-support scale is lower in $51$ of $80$ comparisons and higher in $27$.
In the remaining $2$ comparisons, both endpoints are right censored, so their ordering is unresolved.
At $L=14$, however, the full-support scale is lower in $18$ of $20$ comparisons across generators, parents, and starting signs, and higher in only $2$.
At $L=14$, both panels have median ratios and interquartile ranges below unity.
Finer amplitude sampling retains the full-support advantage in $17$ of $20$ pairs (Appendix~\ref{SecS5_architecture}).

The endpoint scan changes source count, total contrast weight, and placement together, so it does not by itself isolate a spatial effect.
Panels (e) and (f) of Fig.~\ref{Fig_8} make that separation with the compact and dispersed fields defined in Sec.~\ref{Sec2p3_architectures}.
Their response-scale ratios cross unity with size, parent, and probe.
The median dispersed-to-compact ratio is $1.10$ across $75$ spectral comparisons with both onsets observed.
For the $120$ persistent operator comparisons, the median ratio is $1.00$.
The counts of earlier, equal, and later responses are given in Appendix~\ref{SecS5_architecture}.
Placement remains physically visible after matching the source count, the signed content of each brickwall layer, and both first spatial moments, but compact and dispersed patterns have no universal ordering.
This variable ordering differs from the more consistent full-support advantage over a single source.

At fixed homogeneous background and summed squared contrast strength, the largest-size median gap-ratio and operator response scales are higher for full support.
The lower onsets at fixed local amplitude also reflect the larger summed squared contrast strength.
Within this fixed-budget family, scrambling the same set of local gates reduces typical destructive interference and advances the gap-ratio response (Appendix~\ref{SecS5_architecture}).

The distinction between the primary reference $H$ and the sign-matched controls $H_\sigma$ does not generate the endpoint hierarchy.
Each nonuniform circuit is held fixed and compared with both homogeneous references.
In all $80$ comparisons, the choice of reference does not change whether the onset falls within the scanned range.
The median curve correlations are $0.997$ for the spectral response and $1.000$ for the operator response, while the median ratios of the two onset estimates are $1.001$ and $1.000$, respectively.
Individual small-system spectral crossings can shift, so the absolute onset is baseline dependent even though the qualitative architecture hierarchy is not.

At fixed local amplitude, increasing support strongly enhances the finite-size operator susceptibility and biases the largest-size spectral response toward smaller amplitudes, while the parent, generator pair, starting sign, and spatial arrangement control the remaining fluctuations.
The all-$K$ scan covers open half-filled chains, two XXZ parent anisotropies ($\Delta=0.6$ and $1.8$), five fixed generator pairs, and all available odd source counts.

\section{Conclusion}
\label{Sec4_conclusion}

Our central result is that the spatial architecture of a coherent mismatch is an independent control of the finite-size response of an interacting Floquet circuit.
A uniform charge-preserving gate deformation preserves integrability, whereas breaking spatial uniformity at even one bond can reshape the global quasienergy spectrum.
When the mismatch is distributed across the chain, operator dynamics respond at a lower local amplitude than for a single defect, and random-matrix spectral correlations generally appear more readily.
With the homogeneous background and summed squared contrast strength matched, full support instead has higher median gap-ratio and operator onsets at the largest tested sizes.

The response does not vary smoothly with the number of modified bonds.
Patterns with the same source count remain distinguishable after their layer content and first spatial moments are matched, but compact and dispersed arrangements have no universal ordering.
The all-pair perturbative susceptibility at fixed local amplitude attributes the leading separation between a distributed mismatch and a defect to the accumulated weight of many local sources.
Interference provides a smaller correction to the all-pair spectral weight, yet it redistributes nearby-level mixing in a parent- and size-dependent way without consistently favoring the closest levels.
Local observables distinguish the tested spatial patterns, while correlation-hole depth need not track the global spectral response.

Larger systems, broader generator samples, and continuously tunable source patterns are needed to determine which architecture effects survive in the thermodynamic limit.
Their density dependence should be tested at both fixed local and fixed total perturbation strength.
Programmable Floquet processors offer a direct setting in which spatially shaped coherent perturbations can be used to test and control this route from integrable to chaotic dynamics.

\appendix

\section{Bond-resolved Floquet tangent}
\label{SecS1_tangent}
We derive the bond-resolved Floquet tangent and examine how spatial architecture, contrast strength, and diagnostic choices affect the spectral and operator responses.

Consider an even chain of length $L$ with $M=L-1$ bonds.
Let $u_0$ be the parent two-qubit gate, and let $W_{\text{avg}}$ and $W_c$ be the average and contrast generators defined in the main text~\cite{Vanicat2018,Claeys2022,Miao2023}.
We introduce an auxiliary interpolation parameter $\lambda$ and write the gate on bond $b$ as
\begin{equation}
u_b(\delta,\lambda) = \mathrm{e}^{-\mathrm{i}\delta(W_{\text{avg}}+\lambda s_bW_c)}u_0 \,.
\end{equation}
The physical circuit is obtained at $\lambda=1$, while $\lambda=0$ gives the homogeneous reference circuit at the same deformation amplitude $\delta$.
The pattern coefficient $s_b$ specifies the spatial support and sign of the contrast.

For the four nonuniform circuits, the pattern coefficients are
\begin{equation}
\begin{aligned}
s_{b,\text{AB}} &= (-1)^{b-1}, \quad s_{b,\text{BA}} = -(-1)^{b-1} \,,\\
s_{b,D_A} &= \begin{cases}1, & b=b_A,\\0, & b\neq b_A,\end{cases} \quad
s_{b,D_B} = \begin{cases}-1, & b=b_B,\\0, & b\neq b_B,\end{cases} \,,
\end{aligned}
\end{equation}
where $b_A=L/2$ and $b_B=L/2-1$.
Thus AB and BA contain $M$ nonzero sources, whereas each defect circuit contains one.

Order the $M$ gate slots in one Floquet period by $\ell=1,\ldots,M$, and let $b_\ell$ be the bond associated with slot $\ell$.
With $h_\ell(\delta)$ denoting the homogeneous gate in slot $\ell$, the Floquet operator is
\begin{equation}
U_H(\delta) = h_M(\delta)h_{M-1}(\delta)\cdots h_1(\delta) \,.
\end{equation}
The rightmost gate $h_1(\delta)$ acts first, followed by the remaining gates in order.
The local contrast tangent is
\begin{equation}
W_c(\delta) = \int_0^1\mathrm{e}^{-\mathrm{i}x\delta W_{\text{avg}}}W_c\mathrm{e}^{\mathrm{i}x\delta W_{\text{avg}}}\dd{x} \,.
\end{equation}
After embedding this two-qubit operator on bond $b_\ell$, the exact derivative of the local gate at the homogeneous point is
\begin{equation}
\eval{\pdv{u_{b_\ell}(\delta,\lambda)}{\lambda}}_{\lambda=0} = -\mathrm{i}\delta s_{b_\ell}W_{c,b_\ell}(\delta)h_\ell(\delta) \,.
\label{EqS_local_derivative}
\end{equation}

Let $A_\ell(\delta)$ contain all homogeneous gates acting after slot $\ell$:
\begin{equation}
A_\ell(\delta) = h_M(\delta)h_{M-1}(\delta)\cdots h_{\ell+1}(\delta) \,,
\end{equation}
with $A_M(\delta)=\mathbb{1}$.
Differentiating the complete product gives
\begin{equation}
\begin{aligned}
D_{\bm{s}}(\delta) &= \eval{\pdv{U_{\bm{s}}(\delta,\lambda)}{\lambda}}_{\lambda=0} \,,\\
D_{\bm{s}}(\delta) &= -\mathrm{i}\delta K_{\bm{s}}(\delta)U_H(\delta) \,,\\
K_{\bm{s}}(\delta) &= \sum_{\ell=1}^{M}s_{b_\ell}A_\ell(\delta)W_{c,b_\ell}(\delta)A_\ell^\dagger(\delta) \,.
\end{aligned}
\end{equation}
Equivalently, $K_{\bm{s}}(\delta)=\mathrm{i}D_{\bm{s}}(\delta)U_H^\dagger(\delta)/\delta$.
The ordered sum includes coherent interference between insertions on different bonds.

\section{Fixed-sector weights and mixing scale}
\label{SecS2_mixing}

Let $P_N$ project onto the charge sector with $N=L/2$ particles, whose dimension is $d_N=\binom{L}{N}$.
We remove the sector-wide identity component from the tangent according to
\begin{equation}
K_{\bm{s},\circ} = P_NK_{\bm{s}}P_N-\frac{\Tr_N(P_NK_{\bm{s}}P_N)}{d_N}\mathbb{1}_N \,.
\end{equation}
The centered tangent weight is
\begin{equation}
\Gamma_{\bm{s}} = \frac{1}{d_N}\Tr_N[(K_{\bm{s},\circ})^2] \,.
\end{equation}
If $K_{\ell,\circ}$ denotes the centered tangent produced by retaining only slot $\ell$, the one-insertion contribution and its coherent correction are
\begin{equation}
\begin{aligned}
\Gamma_{\bm{s},\text{self}} &= \sum_{\ell=1}^{M}s_{b_\ell}^2\frac{\Tr_N[(K_{\ell,\circ})^2]}{d_N} \,,\\
\Gamma_{\bm{s},\text{cross}} &= \Gamma_{\bm{s}}-\Gamma_{\bm{s},\text{self}} \,.
\end{aligned}
\end{equation}
The cross term vanishes identically for a single nonzero source and tests the coherent addition of distinct insertions in the finite-density circuits.

For spectral mixing, the relevant quantity is the off-diagonal tangent weight in the eigenbasis $\{\ket{\alpha}\}$ of $U_H$, motivated by level-repulsion and random-matrix diagnostics~\cite{Atas2013,Kos2018}.
We define
\begin{equation}
\begin{aligned}
v_{\bm{s},\text{full}} &= \frac{1}{d_N(d_N-1)}\sum_{\alpha\neq\beta}|\mel{\alpha}{K_{\bm{s},\circ}}{\beta}|^2 \,,\\
v_{\bm{s},\text{self}} &= \frac{1}{d_N(d_N-1)}\sum_{\alpha\neq\beta}\sum_{\ell=1}^{M}s_{b_\ell}^2|\mel{\alpha}{K_{\ell,\circ}}{\beta}|^2 \,,\\
v_{\bm{s},\text{cross}} &= v_{\bm{s},\text{full}}-v_{\bm{s},\text{self}} \,.
\end{aligned}
\label{EqS_offdiagonal_variances}
\end{equation}
The self-only variance retains each bond's off-diagonal weight and omits interference between different bonds.

The mean quasienergy spacing in the fixed sector is $2\pi/d_N$.
At first order, a physical contrast amplitude $\delta$ produces a root-mean-square off-diagonal mixing element $\delta\sqrt{v_{\bm{s},q}}$.
Equating this global root-mean-square matrix element with the mean spacing gives the heuristic susceptibility scale
\begin{equation}
\delta_{\text{mix},\bm{s},q} = \frac{2\pi}{d_N\sqrt{v_{\bm{s},q}}}, \quad q\in\{\text{self},\text{full}\} \,.
\end{equation}
For either value of $q$, $\delta_{\text{mix},\text{FD},q}$ is the arithmetic mean over AB and BA, while $\delta_{\text{mix},\text{D},q}$ is the arithmetic mean over $D_A$ and $D_B$.
This estimate compares the mixing susceptibility of different architectures using all off-diagonal tangent elements.
We determine the persistent spectral-response scale separately from the numerical diagnostics.

To resolve nearby levels, write the eigenvalues of $U_H$ as $\mathrm{e}^{-\mathrm{i}\theta_\alpha}$ and define the pairwise full and self-only weights by
\begin{equation}
\begin{aligned}
w_{\alpha\beta,\text{full}} &= |\mel{\alpha}{K_{\bm{s},\circ}}{\beta}|^2 \\
w_{\alpha\beta,\text{self}} &= \sum_{\ell=1}^M s_{b_\ell}^2|\mel{\alpha}{K_{\ell,\circ}}{\beta}|^2 \,.
\end{aligned}
\end{equation}
Let $P_{\text{adj}}$ contain the cyclically adjacent phase pairs and let $P_{10}$ contain pairs whose circular phase distance does not exceed $0.1\pi$.
For $q\in\{\text{full},\text{self}\}$ and either pair set $P$, we use
\begin{equation}
v_{\bm{s},q,P} = \frac{1}{|P|}\sum_{(\alpha,\beta)\in P}w_{\alpha\beta,q} \,.
\end{equation}
We also define the spacing-weighted susceptibility
\begin{equation}
\chi_{\bm{s},q}(\eta) = \frac{1}{d_N}\sum_{\alpha\neq\beta}\frac{w_{\alpha\beta,q}}{4\sin^2[(\theta_\alpha-\theta_\beta)/2]+\eta^2}, \quad \eta = \frac{2\pi}{d_N} \,.
\end{equation}
For each of these selections, the cross fraction is the difference between the full and self-only values divided by the full value.

To expose the support dependence, let $\kappa^2$ denote the approximately position-independent off-diagonal variance of one bulk source.
The self-only variances then satisfy
\begin{equation}
v_{\text{FD},\text{self}} \simeq M\kappa^2, \quad v_{\text{D},\text{self}} \simeq \kappa^2 \,,
\end{equation}
and hence
\begin{equation}
\frac{\delta_{\text{mix},\text{FD},\text{self}}}{\delta_{\text{mix},\text{D},\text{self}}} \simeq \frac{1}{\sqrt{M}} \,.
\end{equation}
This inverse-square-root law follows from adding the self-only weights of $M$ comparable sources.

The correction produced by coherent interference can be expressed through the finite-density cross fraction
\begin{equation}
f_{\text{cross}} = \frac{v_{\text{FD},\text{cross}}}{v_{\text{FD},\text{full}}} \,.
\end{equation}
The full and self-only mixing scales then obey
\begin{equation}
\frac{\delta_{\text{mix},\text{FD},\text{full}}}{\delta_{\text{mix},\text{FD},\text{self}}} = \sqrt{1-f_{\text{cross}}} \,.
\label{EqS_cross_correction}
\end{equation}
A positive $f_{\text{cross}}$ lowers the finite-density mixing scale but does not change the source-count exponent unless it develops an additional systematic size dependence.

\section{Numerical implementation and tangent extensivity}
\label{SecS3_implementation}

We evaluate Eq.~\eqref{EqS_local_derivative} using the Fr\'echet derivative of the matrix exponential and check it against a centered finite difference.
Unless stated otherwise, we use half-filled chains of length $L=8$, $10$, $12$, and $14$.
The parent anisotropies are $\Delta=0.6$ and $1.8$, with reference amplitude $\delta_{\text{ref}}=0.04$.
The off-diagonal matrix elements are evaluated using the orthonormal complex-Schur vectors of the homogeneous Floquet operator.
The spectral scale $\delta_{\text{spec}}$ requires the cyclic gap-ratio response, full-spacing preference, and circular-rigidity gain to meet their main-text thresholds simultaneously.
We take the first of two consecutive sampled amplitudes satisfying this condition.
Right-censored cases are shown as bounds and excluded from correlations and error measures.

\begin{figure*}[htbp!]
\centering
\includegraphics[width=\linewidth]{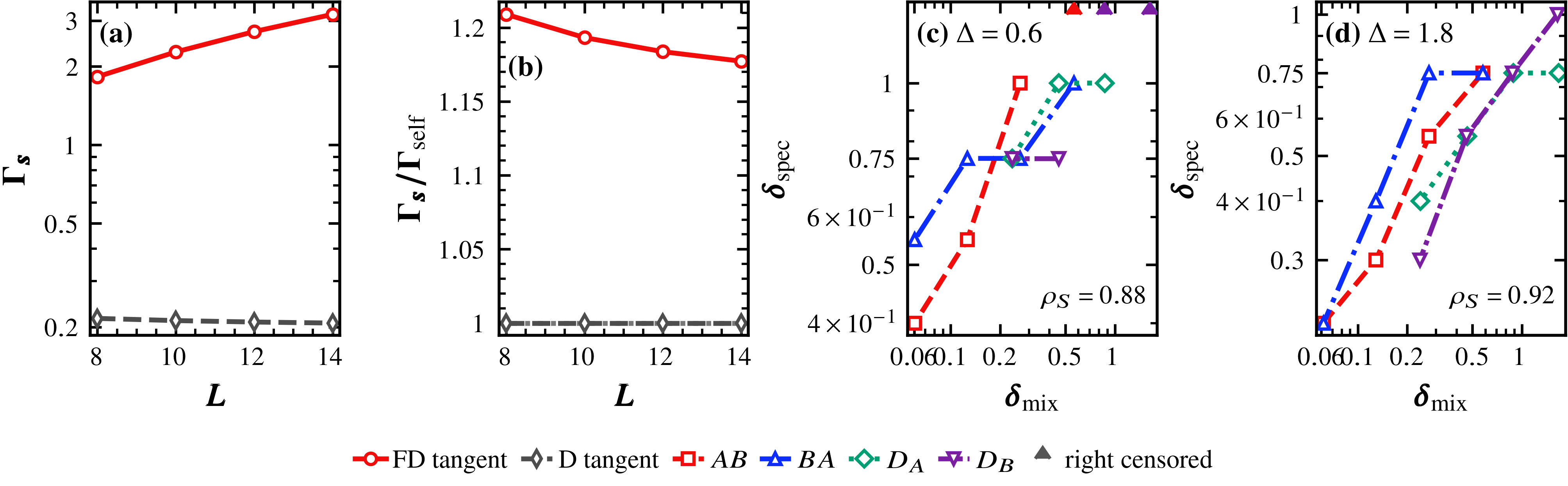}
\caption{Bond-resolved tangent weights and spectral mixing estimates for finite-density and single-defect deformations.
Panel (a) shows the centered tangent weight $\Gamma_{\bm{s}}$ at $\delta_{\text{ref}}=0.04$.
The finite-density tangent is extensive in $L$, whereas the single-defect tangent remains of order one.
Panel (b) shows the ratio of the total tangent weight to the self-only contribution.
The finite-density ratio exceeds unity because of constructive cross terms between spatially separated insertions, while a single defect has no cross term.
The curves for $\Delta=0.6$ and $1.8$ overlap in panels (a) and (b), so only one curve is shown for each mismatch extent.
Panels (c) and (d) compare the measured spectral crossover scale $\delta_{\text{spec}}$ with the tangent estimate $\delta_{\text{mix}}$ for $\Delta=0.6$ and $1.8$, respectively.
Upward triangles mark right-censored values of $\delta_{\text{spec}}$.
The Spearman correlations over uncensored points are $\rho_S=0.88$ and $0.92$.}
\label{Fig_9}
\end{figure*}

Figure~\ref{Fig_9}(a) shows that the full finite-density tangent weight grows linearly with system size.
Over the tested sizes, a linear fit gives $\Gamma_{\bm{s}}\simeq0.224L+0.0346$.
Over the same sizes, the single-defect weight decreases only weakly from $0.2159$ to $0.2072$ and remains of order one.

The finite-density ratio $\Gamma_{\bm{s}}/\Gamma_{\bm{s},\text{self}}$ lies between $1.177$ and $1.209$ in Fig.~\ref{Fig_9}(b).
The corresponding cross fraction of the total centered tangent weight is $15.1\%$-$17.3\%$.
For this generator pair, the total weight is independent of the parent, linking its extensivity to the local generators and their spatial support.
The off-diagonal weight, however, depends on the eigenbasis of the homogeneous Floquet operator.

Figures~\ref{Fig_9}(c) and~\ref{Fig_9}(d) compare that off-diagonal projection with the measured persistent spectral-response scale.
The finite-density mixing estimate is $0.256$-$0.348$ of the corresponding defect estimate over the complete grid.
The tangent estimate gives Spearman coefficients of $0.88$ for $\Delta=0.6$ and $0.917$ for $\Delta=1.8$ across system sizes and mismatch extents.

\section{Source-count null test and reference-amplitude robustness}
\label{SecS4_null}

We isolate the role of interference by comparing the full variance $v_{\bm{s},\mathrm{full}}$ with the self-only variance $v_{\bm{s},\mathrm{self}}$ in Eq.~\eqref{EqS_offdiagonal_variances}.
Retaining each bond's exact weight preserves boundary effects, layer ordering, and unequal source strengths in the self-only baseline.

\begin{figure*}[htbp!]
\centering
\includegraphics[width=\linewidth]{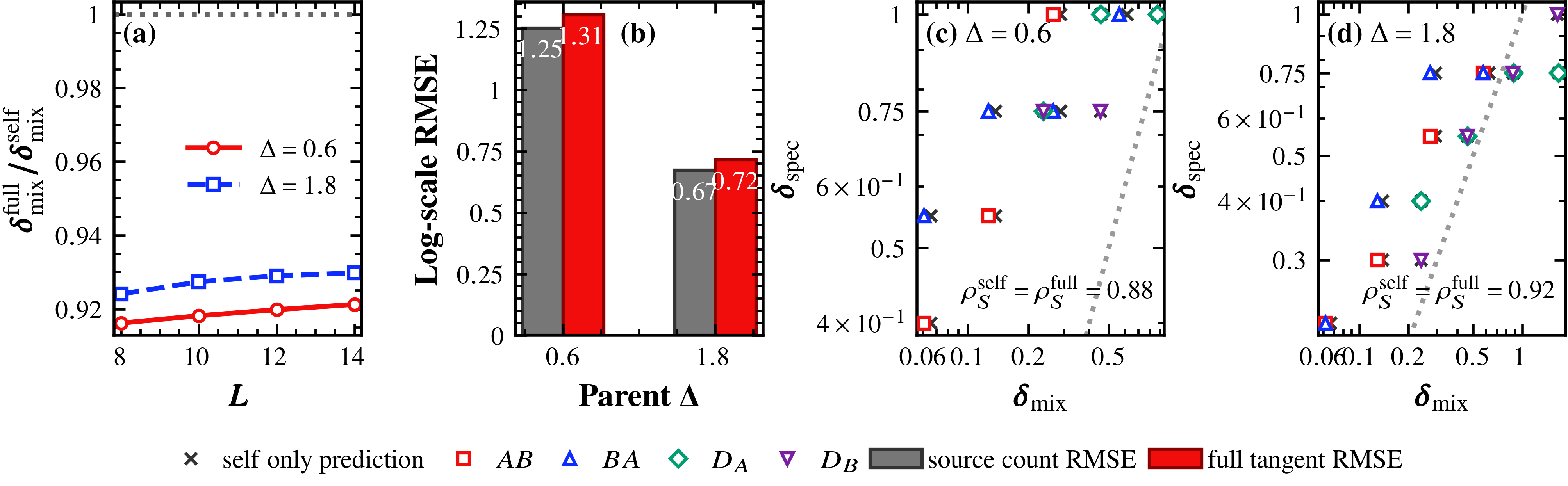}
\caption{Comparison of self-only and full tangent estimates at $\delta_{\text{ref}}=0.04$.
Panel (a) shows the ratio of the full tangent prediction $\delta_{\text{mix},\text{full}}$ to the self-only prediction $\delta_{\text{mix},\text{self}}$ for the finite-density circuits.
Panel (b) shows the log-scale root-mean-square error of the source-count and full tangent predictions relative to the observed $\delta_{\text{spec}}$, evaluated over uncensored cases.
Panels (c) and (d) show the observed scale against the two predictions for $\Delta=0.6$ and $1.8$, respectively.
Black crosses denote the source-count prediction, colored open markers denote the full tangent prediction, and short horizontal segments connect the two estimates for the same circuit.
The dotted line denotes equality between the predicted and observed scales.
The two models give identical Spearman rankings, while the self-only model has a slightly smaller log-scale error for both parents.}
\label{Fig_10}
\end{figure*}

The finite-density off-diagonal cross fraction lies between $0.1355$ and $0.1606$.
For the finite-density circuit, Eq.~\eqref{EqS_cross_correction} gives a full-to-self mixing-scale ratio between $0.916$ and $0.93$, consistent with Fig.~\ref{Fig_10}(a).
Coherent interference is constructive and lowers the predicted finite-density scale by approximately $7\%$-$9\%$.
This correction is smaller than the separation between finite-density and defect estimates.

The source-count and full tangent estimates give identical Spearman and Kendall rankings for both parent circuits.
For $\Delta=0.6$, the Spearman coefficient is $0.88$ and the Kendall coefficient is $0.792$.
For $\Delta=1.8$, the corresponding values are $0.917$ and $0.831$.
The log-scale root-mean-square errors of the source-count and full tangent estimates are $1.252$ and $1.306$ for $\Delta=0.6$, and $0.672$ and $0.715$ for $\Delta=1.8$, respectively.

We repeated the bond-resolved calculation at $\delta_{\text{ref}}=0.02$, $0.04$, and $0.08$.
Across both parents and all four sizes, the self-only scaling ratio satisfies:
\begin{equation}
0.9951 \leq \sqrt{M}\frac{\delta_{\text{mix},\text{FD},\text{self}}}{\delta_{\text{mix},\text{D},\text{self}}} \leq 1.0011 \,.
\end{equation}
Including the cross terms gives
\begin{equation}
0.9144 \leq \sqrt{M}\frac{\delta_{\text{mix},\text{FD},\text{full}}}{\delta_{\text{mix},\text{D},\text{full}}} \leq 0.9312 \,.
\end{equation}
The Spearman rankings remain unchanged throughout this reference-amplitude scan.
The scaling collapse shown in Fig.~\ref{Fig_7} of the main text is insensitive to the selected linearization point over the tested range.

\begin{figure*}[htbp!]
\centering
\includegraphics[width=\linewidth]{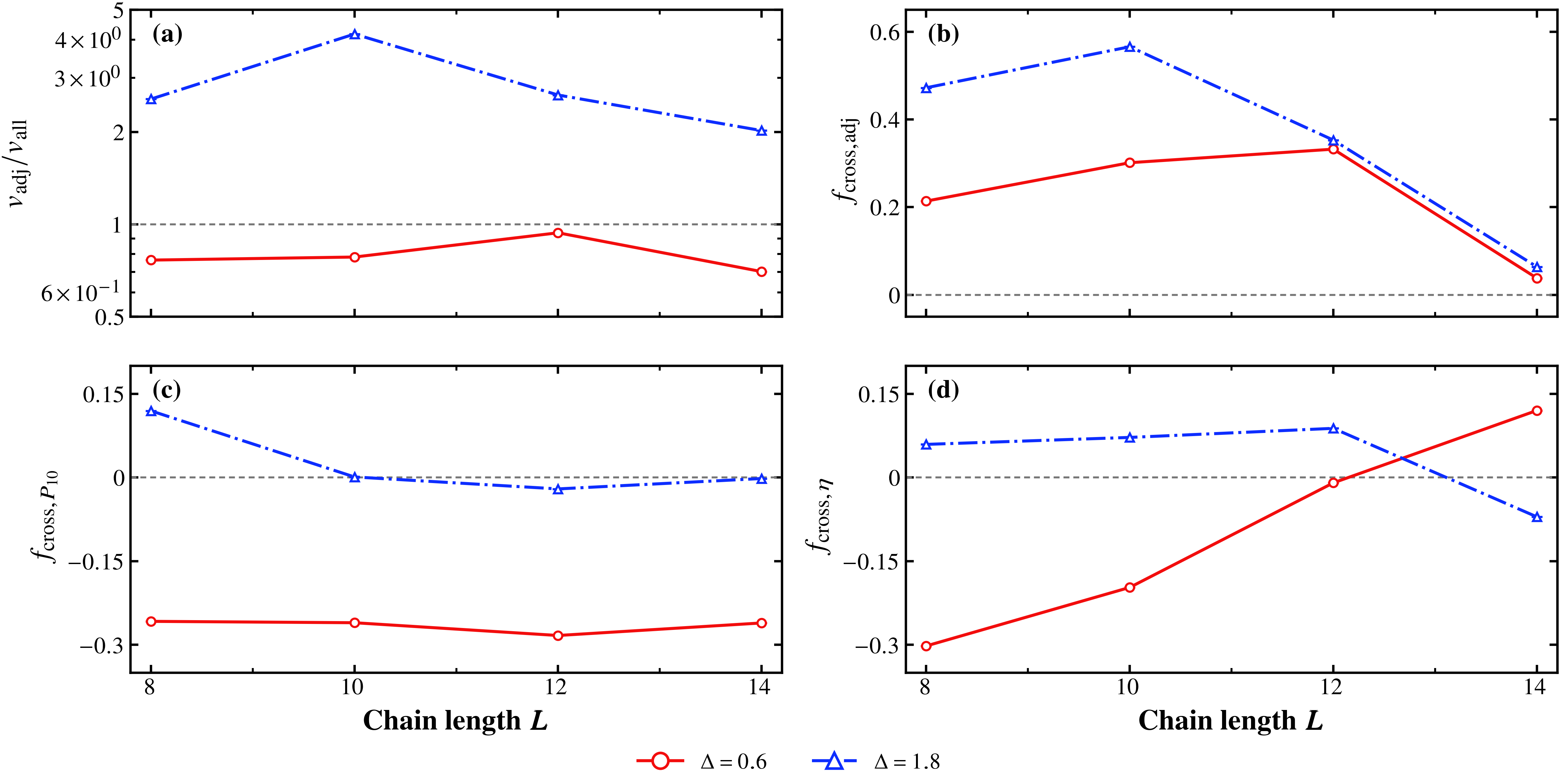}
\caption{Near-level matrix elements of the finite-density tangent at $\delta_{\text{ref}}=0.04$.
The plotted values are arithmetic means over the two alternating patterns, $AB$ and $BA$.
Panel (a) compares the mean tangent weight of cyclically adjacent eigenphases with the all-pair mean.
Panel (b) gives the adjacent-pair cross fraction.
Panel (c) gives the cross fraction for $P_{10}$, the set of pairs whose circular phase distance does not exceed $0.1\pi$.
Panel (d) gives the spacing-weighted cross fraction at $\eta=2\pi/d_N$.
Red circles and blue triangles denote $\Delta=0.6$ and $1.8$, respectively.
{Adjacent-level mean weights lie below the all-pair mean at $\Delta=0.6$ and above it at $\Delta=1.8$.
The cross fractions in (c,d) take both signs, showing that interference can enhance or suppress the weight depending on the level selection and weighting.}}
\label{Fig_11}
\end{figure*}

Figure~\ref{Fig_11}(a) shows that the adjacent-level tangent weight depends strongly on the parent circuit.
The adjacent-to-all-pair ratio lies between $0.7$ and $0.94$ for $\Delta=0.6$, but between $2.02$ and $4.17$ for $\Delta=1.8$.
The adjacent-pair cross fraction is positive throughout the grid and ranges from $0.038$ to $0.566$.
This positive sign does not persist under other near-level selections.
The cross fraction for phase pairs whose circular phase distance does not exceed $0.1\pi$ ranges from $-0.283$ to $0.119$, and the spacing-weighted fraction ranges from $-0.302$ to $0.12$.
The Spearman coefficient between the logarithms of the adjacent gap and adjacent coupling strength lies between $-0.005$ and $0.106$.
The sign and magnitude of the near-level interference contribution depend on the parent, system size, and level selection.

\section{Architecture scan and diagnostic sensitivity}
\label{SecS5_architecture}

We extend the main-text comparison to five fixed generator pairs, $G_0$-$G_4$, with $G_0$ denoting the original pair.
The additional pairs are drawn in the Frobenius-orthonormal basis
\begin{equation}
B = (Z_L/2,Z_R/2,Z_LZ_R/2,T/\sqrt{2},J/\sqrt{2}) \,.
\end{equation}
For $g=1,\ldots,4$ and $\mu\in\{A,B\}$, the generators are
\begin{equation}
W_{\mu,g} = \sum_{a=1}^5 c_{\mu a,g}B_a \,.
\end{equation}
For each generator pair, we draw two independent five-component vectors with independent standard-normal entries.
Normalizing each vector to unit Euclidean norm gives $\norm{W_{\mu,g}}_F=1$.
We retain pairs satisfying $|\bm{c}_{A,g}\cdot\bm{c}_{B,g}|=|\Tr(W_{A,g}W_{B,g})|\leq0.65$, discarding both vectors otherwise.
Table~\ref{TabS_generators} lists the first four accepted pairs, selected independently of the circuit, spectral, and operator-response data.

\begin{table*}[htbp!]
\centering
\caption{Coefficients of the four additional generator pairs.
Each row specifies one unit-Frobenius generator in the basis $B$.
Coefficients are shown to four decimal places; all calculations use the unrounded values.}
\begingroup
\setlength{\tabcolsep}{4pt}
\begin{ruledtabular}
\begin{tabular}{ccrrrrr}
Pair & Generator & $c_{Z_L}$ & $c_{Z_R}$ & $c_{Z_LZ_R}$ & $c_T$ & $c_J$ \\
\hline
$G_1$ & $W_A$ & $0.641$ & $0.7386$ & $0.0358$ & $0.001$ & $0.2058$ \\
$G_1$ & $W_B$ & $0.2251$ & $-0.4591$ & $0.0846$ & $-0.4366$ & $-0.7354$ \\
$G_2$ & $W_A$ & $0.558$ & $-0.1286$ & $-0.6657$ & $-0.3841$ & $0.2853$ \\
$G_2$ & $W_B$ & $-0.0988$ & $0.0246$ & $0.2495$ & $-0.7179$ & $0.6419$ \\
$G_3$ & $W_A$ & $-0.3448$ & $0.2912$ & $0.5209$ & $-0.4027$ & $0.6023$ \\
$G_3$ & $W_B$ & $0.302$ & $-0.7606$ & $0.303$ & $-0.2214$ & $0.4353$ \\
$G_4$ & $W_A$ & $0.1164$ & $-0.4636$ & $0.0546$ & $0.8759$ & $-0.0366$ \\
$G_4$ & $W_B$ & $0.2093$ & $-0.7803$ & $0.1004$ & $-0.3266$ & $0.4802$ \\
\end{tabular}
\end{ruledtabular}
\endgroup
\label{TabS_generators}
\end{table*}

For each generator pair, parent, size, and starting sign, the scan includes every odd source count $K=1,3,\ldots,M$.
We number the bonds from $1$ to $M$ and denote the ordered set of active bonds by $\mathcal{A}_K=\{b_0,\ldots,b_{K-1}\}$.
The starting sign of the source pattern is denoted by $\sigma\in\{+1,-1\}$.
The single-source endpoint has $b_0=(M+1)/2$, while the clustered support uses $b_{i,\text{cl}}=(M-K)/2+1+i$.
For $K>1$, the dispersed support uses $b_{i,\text{disp}}=1+\lfloor i(M-1)/(K-1)+1/2\rfloor$, and at $K=M$ every bond is active.
The signs alternate according to active-source rank, including across inactive gaps:
\begin{equation}
\begin{aligned}
s_{b_i,K,\sigma} &= \sigma(-1)^i, \quad i=0,\ldots,K-1,\\
s_{b,K,\sigma} &= 0 \text{ for } b\notin\mathcal{A}_K \,.
\end{aligned}
\end{equation}
Because the source count $K$ is odd, the signed sums satisfy
\begin{equation}
\sum_{b=1}^M s_{b,K,\sigma} = \sigma, \quad \frac{1}{M}\sum_{b=1}^M W_b = W_{\text{avg}}+\frac{\sigma}{M}W_c \,.
\end{equation}
Each architecture is compared with a homogeneous control using $W_{\text{ctrl},\sigma}=W_{\text{avg}}+\sigma W_c/M$ on every bond.
The original five-circuit comparison uses the shared $W_{\text{avg}}$ reference, while the all-$K$ scan uses separate homogeneous controls for the two starting signs.
Here $K$ counts the nonzero entries of the raw contrast field $s_b$.
Relative to the sign-matched homogeneous control, the centered contrast field is
\begin{equation}
\tilde{s}_{b,K,\sigma} = s_{b,K,\sigma}-\sigma/M \,.
\end{equation}
Bonds that are inactive in the raw architecture carry the compensating background $-\sigma/M$ relative to the matched control.
Thus $K$ counts the original sources, rather than all bonds that differ from the homogeneous reference.
We keep the per-source contrast amplitude and signed generator mean fixed as $K$ and the spatial arrangement vary.
The source count, support, total contrast weight, and momentum content can therefore change.
Matching the mean generators does not require equal arithmetic averages of the gates after exponentiation at finite $\delta$.

To isolate placement, we choose the odd source count $K$ nearest $M/2$ at each size, taking the smaller value in a tie.
Within each brickwall layer, the compact and dispersed fields have identical active-source counts and signed sums.
Their active and signed first spatial moments are also identical across the full chain.
Among fields satisfying these constraints, we select the pair with the largest difference in the support-spread measure
\begin{equation}
S_p = \sum_{b<b'}s_{b,p}^2s_{b',p}^2(b-b')^2, \quad p\in\{\text{comp},\text{disp}\} \,.
\end{equation}
The resulting positive-sign fields are listed in Table~\ref{TabS_layer_matched_patterns}, and the negative-sign fields follow by reversing every sign.

\begin{table*}[htbp!]
\centering
\caption{Placement pairs with matched layer counts and spatial moments, with bonds numbered from $1$ to $M$.
Each ordered pair gives the bond index and source sign for the positive-sign field.
All unlisted bonds use $u_{\text{avg}}$.}
\begin{ruledtabular}
\begin{tabular}{cccc}
$L$ & $K$ & Compact field & Dispersed field \\
\hline
$8$ & $3$ & $\{(3,+),(4,-),(5,+)\}$ & $\{(1,+),(4,-),(7,+)\}$ \\
$10$ & $5$ & $\{(3,-),(4,+),(5,+),(6,+),(7,-)\}$ & $\{(1,-),(2,+),(5,+),(8,+),(9,-)\}$ \\
$12$ & $5$ & $\{(4,-),(5,+),(6,+),(7,+),(8,-)\}$ & $\{(1,+),(2,-),(6,+),(10,-),(11,+)\}$ \\
$14$ & $7$ & $\{(4,-),(5,+),(6,+),(7,-),(8,+),(9,+),(10,-)\}$ & $\{(1,+),(2,+),(4,-),(6,-),(11,-),(12,+),(13,+)\}$ \\
\end{tabular}
\end{ruledtabular}
\label{TabS_layer_matched_patterns}
\end{table*}

\begin{figure*}[htbp!]
\centering
\includegraphics[width=\linewidth]{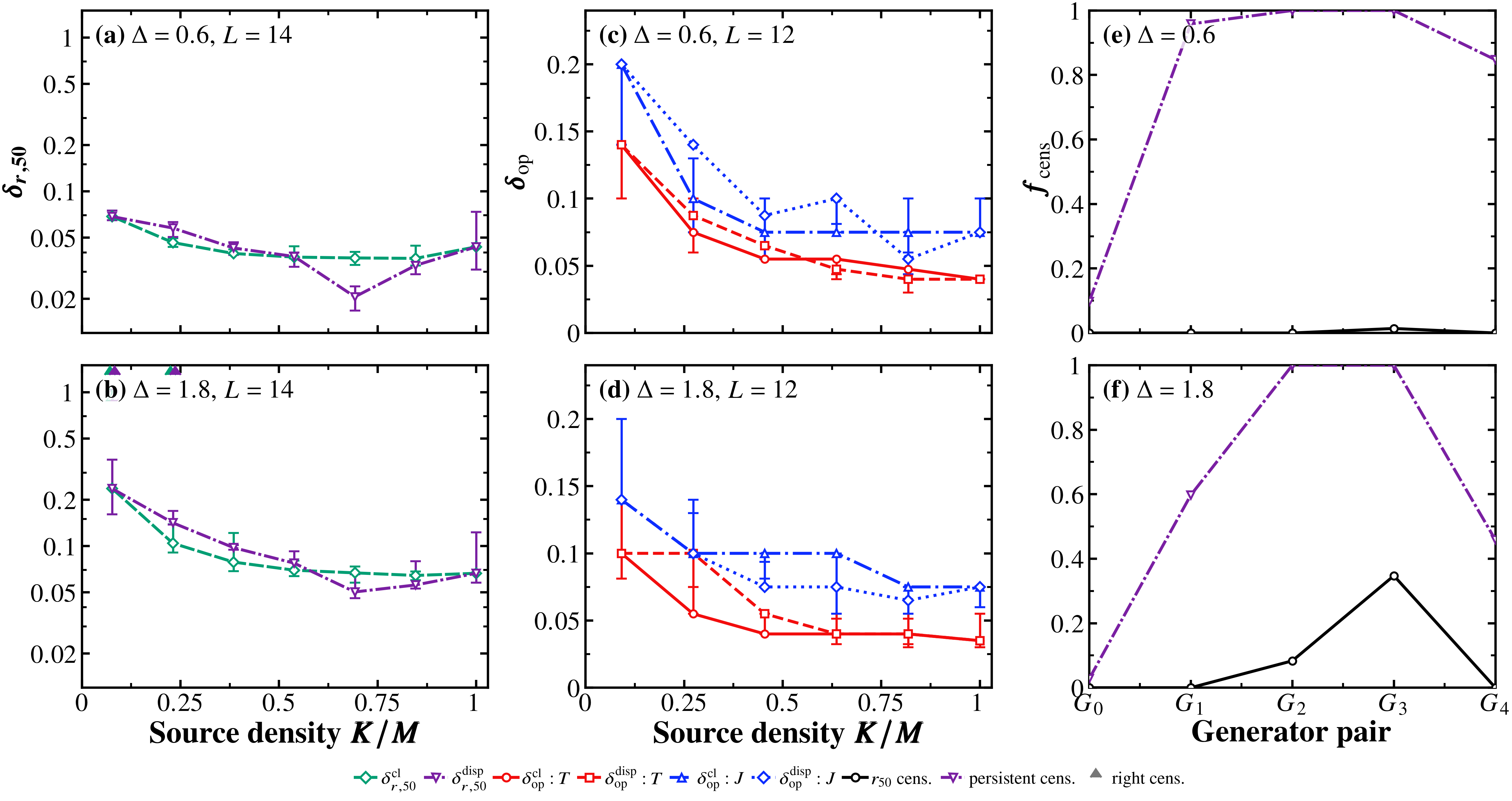}
\caption{Source-count dependence of response scales and fractions of unresolved onsets.
Panels (a) and (b) show the interpolated spectral scale $\delta_{r,50}$ against source fraction $K/M$ at $L=14$ for $\Delta=0.6$ and $1.8$, respectively.
Green diamonds and purple triangles denote clustered and dispersed intermediate-source arrangements.
Observed points and error bars give the median and interquartile range over the five fixed generator pairs and both starting signs.
The two curves share the $K=1$ and $K=M$ endpoint data.
Upward carets mark right-censored $r_{50}$ cases at the spectral scan boundary $\delta=1.3$.
Panels (c) and (d) show the persistent absolute operator-response scale at $L=12$ and $t_{\text{max}}=48$ for the probes $T$ and $J$.
Solid or dash-dotted curves denote clustered arrangements, while dashed or dotted curves denote dispersed arrangements.
Panels (e) and (f) show the right-censored fraction for $\delta_{r,50}$ and for the persistent joint spectral criterion.
For each generator and parent, the censoring fraction includes all $72$ combinations of architecture, system size, and starting sign.
{Median response scales are lower at full support than at a single source, with arrangement-dependent nonmonotonicity at intermediate counts.
The joint spectral criterion remains unresolved in many cases, motivating the broader comparison using $\delta_{r,50}$.}}
\label{Fig_12}
\end{figure*}

Figures~\ref{Fig_12}(a) and~\ref{Fig_12}(b) show that the median spectral scale generally decreases as the source support grows, but the individual source-count paths are not strictly monotonic.
The clustered and dispersed medians also separate at several intermediate source counts.
These curves establish the endpoint support trend but do not isolate placement, because their brickwall-layer content can differ at intermediate $K$.

Full support responds earlier than a single source more consistently in the operator diagnostics [Figs.~\ref{Fig_12}(c,d)].
This comparison includes both parents, three operator sizes, both signs, five generator pairs, and two probes.
The intermediate-$K$ responses vary nonmonotonically with source count and depend on the arrangement.

The controlled placement pairs in Table~\ref{TabS_layer_matched_patterns} separate those fluctuations from source count, layer composition, and linear spatial bias.
For the spectral scale, both compact and dispersed onsets are observed in $75$ of $80$ comparisons across generators, parents, system sizes, and starting signs.
The dispersed field responds at a lower amplitude in $30$ cases and at a higher amplitude in $45$.
The median onset ratio is $1.1$, and the median symmetric absolute difference is $31.7\%$.
For the persistent absolute operator scale at $t_{\text{max}}=48$, all $120$ comparisons are observed.
The dispersed field responds first in $24$ cases, at the same sampled amplitude in $40$, and later in $56$.
The median ratio is $1$, and the median symmetric absolute difference is $30.8\%$.
Whether the compact or dispersed arrangement responds first depends on system size, parent, generator, and probe.

Figures~\ref{Fig_12}(e) and~\ref{Fig_12}(f) show how often the two spectral criteria remain unresolved.
The $r_{50}$ scale is observed for nearly all generators and architectures, whereas the persistent joint criterion is right censored for most cases associated with $G_1$-$G_4$.
The original pair $G_0$ is atypically favorable for the strict persistent criterion, especially at $\Delta=0.6$.
No censored value is replaced by the scan boundary in the medians, endpoint ratios, or ordering probabilities.

We check the antiunitary symmetry at $L=8$ for five generator pairs, two parents, and four amplitudes ($\delta=0.02,0.10,0.40,1.00$).
The comparison includes fourteen patterns, among them the matched homogeneous controls.
All {$560$ tested} cases satisfy the antiunitary symmetry relation and belong to the same symmetry class.
The main-text checks for integrable structure concern the four nonuniform circuits $AB$, $BA$, $D_A$, and $D_B$ constructed from the original generator pair $G_0$.

\begin{figure*}[htbp!]
\centering
\includegraphics[width=\linewidth]{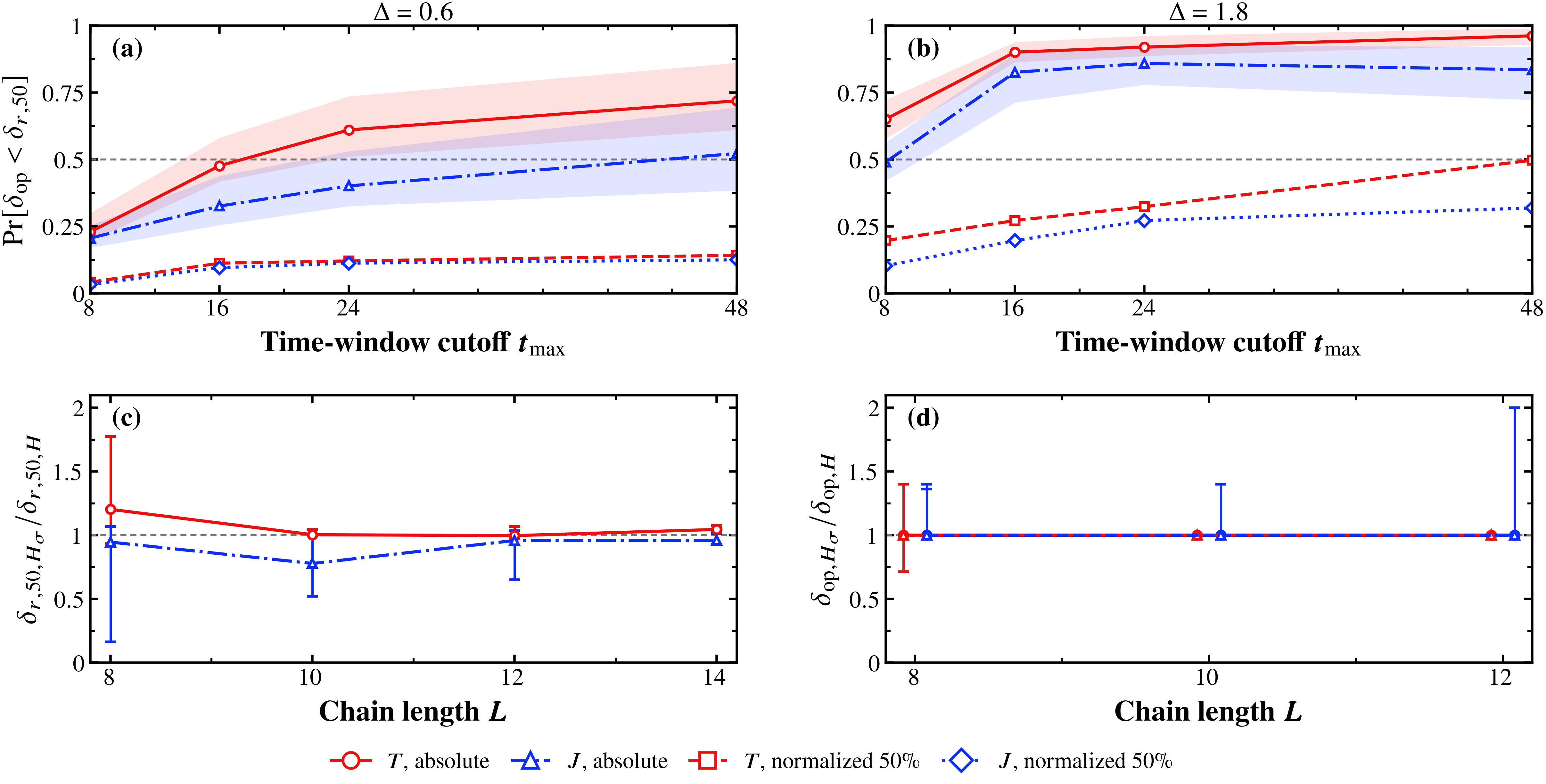}
\caption{Dependence of response ordering on the diagnostic criterion and homogeneous reference.
Panels (a) and (b) show the fraction of observed architecture cases for which $\delta_{\text{op}}<\delta_{r,50}$ as the time-window cutoff changes.
Red circles and blue triangles show the persistent absolute criterion for $T$ and $J$, while the shaded regions span threshold multipliers from $0.75$ to $1.25$.
Red squares and blue diamonds show the corresponding normalized $50\%$ criterion.
Panels (c) and (d) compare onsets for the same nonuniform circuits using either the sign-matched reference $H_\sigma$ or the shared reference $H$.
Panel (c) shows the spectral scale, with red circles and blue triangles denoting $\Delta=0.6$ and $1.8$.
Panel (d) shows the operator scale, with red and blue denoting $T$ and $J$, circles and solid lines denoting $\Delta=0.6$, and triangles and dashed lines denoting $\Delta=1.8$.
Points and bars in panels (c) and (d) give the median and range over the four nonuniform circuits.
{The fraction of cases with an operator onset below the spectral onset depends strongly on the response definition and observation window.
Changing the homogeneous baseline leaves median onset ratios near unity, although individual spectral onsets can shift.}}
\label{Fig_13}
\end{figure*}

Figures~\ref{Fig_13}(a) and~\ref{Fig_13}(b) compare the ordering of operator and spectral onsets under absolute and normalized operator-response criteria.
At $t_{\text{max}}=48$, the original absolute criterion gives $\delta_{\text{op}}<\delta_{r,50}$ in $83\%$ of the observed $T$ comparisons and $67\%$ of the observed $J$ comparisons.
The corresponding fractions are $31\%$ and $22\%$ for the normalized $50\%$ criterion.
With the absolute criterion, operator onsets usually precede $\delta_{r,50}$; normalization changes that ordering in many cases.

Figures~\ref{Fig_13}(c) and~\ref{Fig_13}(d) compare onsets for the same four nonuniform circuits using the two homogeneous references.
In all $80$ comparisons, the choice of reference does not change whether the onset falls within the scanned range.
Across the $32$ spectral cases, the median Spearman correlation between response curves is $0.997$.
The median onset ratio using $H_\sigma$ relative to $H$ is $1.001$.
For the $48$ operator cases, the median curve coefficient and onset ratio are both $1$, with $42$ onsets unchanged on the sampled grid.
The qualitative hierarchy is stable across the two homogeneous baselines, while individual small-system spectral onsets can shift.

\begin{figure*}[htbp]
\centering
\includegraphics[width=6.7in]{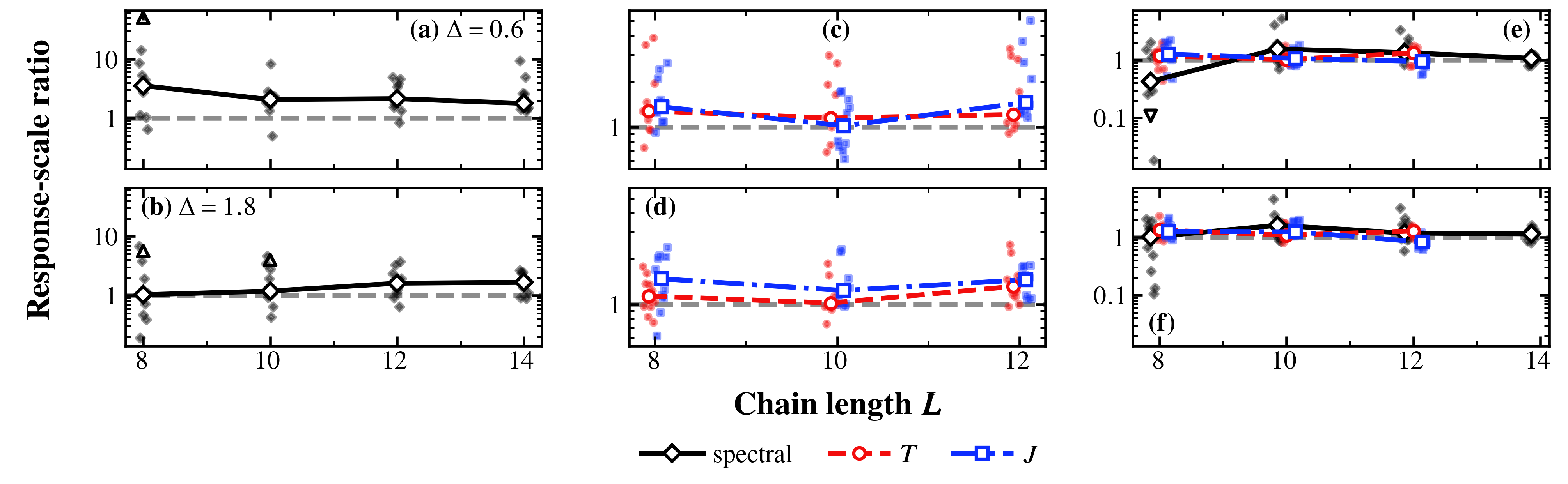}
\caption{{Spatial response at matched contrast budget and homogeneous background $a=0.04$.
Upper and lower rows correspond to $\Delta=0.6$ and $\Delta=1.8$, respectively.
Panels (a) and (b) show the full-support-to-single-source ratio of $\epsilon_{r,50}$ for the five generator pairs $G_0$-$G_4$ and both starting signs.
Panels (c) and (d) give the corresponding ratios of $\epsilon_{\text{op}}$ for $T$ and $J$ at $t_{\text{max}}=48$.
Panels (e) and (f) compare the dispersed and compact fields of Table~\ref{TabS_layer_matched_patterns} at the same source count and budget.
Small points show individual paired ratios, vertical bars show bounds from the sampled onset intervals, and connected large points show medians over pairs with both onsets observed.
Open upward and downward triangles indicate lower and upper bounds when an onset lies beyond the scanned range.
The horizontal line marks equal response scales; connecting segments are guides to the eye.
At matched budget, full support generally has a higher median onset than a single source.
Dispersed arrangements can respond either earlier or later than compact ones.}}
\label{Fig_14}
\end{figure*}

\begin{figure*}[htbp]
\centering
\includegraphics[width=6.7in]{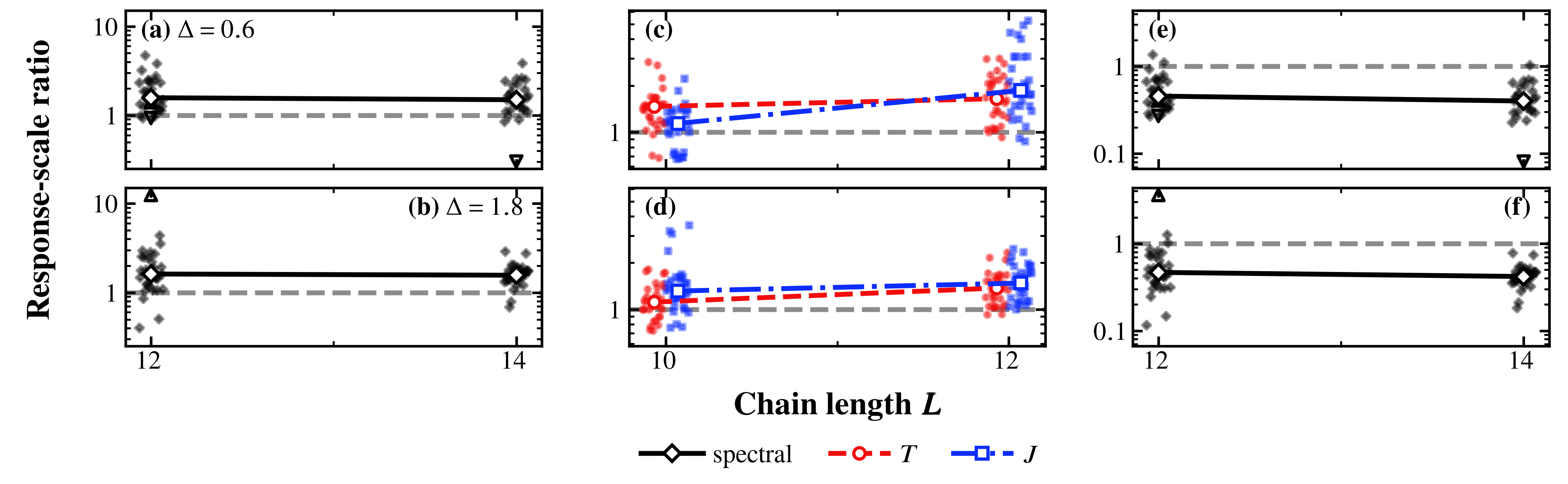}
\caption{{Endpoint response across fifteen additional generator pairs at $a=0.04$.
Upper and lower rows correspond to $\Delta=0.6$ and $\Delta=1.8$, respectively.
Panels (a) and (b) show the full-support-to-single-source ratio of the equal-budget spectral scale $\epsilon_{r,50}$.
Panels (c) and (d) show the corresponding persistent operator-scale ratios at $t_{\text{max}}=48$.
Panels (e) and (f) express the spectral onsets in the local source amplitude $\eta$, related to $\epsilon$ by Eq.~\eqref{EqS_budget_local_amplitude}; they describe the same circuits under a different amplitude normalization.
Points, interval bars, median lines and one-sided bounds have the meanings given in Fig.~\ref{Fig_14}.
The thirty comparisons per parent and size come from fifteen generator pairs, each evaluated with both starting signs.
Full support typically responds later than a single source at equal budget, but earlier at equal local amplitude.}}
\label{Fig_15}
\end{figure*}

The fixed-local-amplitude comparisons above combine spatial extent with an increase in total contrast strength.
To separate these effects, we hold the homogeneous background fixed and redistribute a fixed contrast budget among the bonds.
For each raw field $s_b$, define $\bar{s}=M^{-1}\sum_b s_b$ and the normalized contrast direction $\widehat{W}_c=W_c/\norm{W_c}_F$.
The corresponding gates are
\begin{equation}
\begin{aligned}
q_b &= \frac{s_b-\bar{s}}{\sqrt{\sum_{b'=1}^{M}(s_{b'}-\bar{s})^2}}, \\
u_b(a,\epsilon) &= \mathrm{e}^{-\mathrm{i}V_b}u_0, \\
V_b &= aW_{\text{avg}}+\epsilon q_b\widehat{W}_c \,.
\end{aligned}
\label{EqS_budget_gates}
\end{equation}
Thus the bond average $\bar{V}=M^{-1}\sum_b V_b$ and the total contrast budget satisfy:
\begin{equation}
\bar{V}=aW_{\text{avg}},
\quad
\mathcal{B}=\sum_{b=1}^{M}\norm{V_b-\bar{V}}_F^2=\epsilon^2 \,.
\label{EqS_budget_norm}
\end{equation}
The homogeneous reference has $\epsilon=0$, and $a=0.04$ is fixed throughout Figs.~\ref{Fig_14} and~\ref{Fig_15}.
This budget measures the spatially summed strength of the local contrast generators, rather than a distance between the complete many-body Floquet operators.
As in the sign-matched construction above, the centered single-source field includes a uniform compensating background; its raw source count remains $K=1$.

We denote the gap-ratio envelope's $50\%$ response scale by $\epsilon_{r,50}$ and the persistent absolute operator scale by $\epsilon_{\text{op}}$.
The latter uses the same OSEE and OTOC thresholds, $0.075$ bits and $0.020$, over $t_{\text{max}}=48$.
At matched contrast budget, full support typically responds later than a single source [Fig.~\ref{Fig_14}].
At $L=14$, full support reaches the gap-ratio onset later in all ten combinations of generator and starting sign for $\Delta=0.6$.
For $\Delta=1.8$, it responds later in eight of the ten combinations.
The median full-support-to-single-source ratios are $1.78$ and $1.67$, respectively.
Both neutral probes also have median operator-scale ratios above unity at $L=12$.
The lower full-support onsets at fixed local amplitude therefore reflect, in part, the larger total contrast strength.

Position still affects the response when the budget and raw source count are both fixed.
The compact and dispersed fields retain the matched layer content and first spatial moments of Table~\ref{TabS_layer_matched_patterns}.
At $L=14$, the gap-ratio onset-ratio intervals exclude unity in eighteen of twenty comparisons across generators, parents, and starting signs.
The dispersed field responds earlier in five of these pairs and later in thirteen.
For example, the positive-sign $G_0$ field at $\Delta=0.6$ has a dispersed-to-compact ratio of $0.790$, bounded by $[0.760,0.820]$.
The positive-sign $G_2$ field at $\Delta=1.8$ instead gives $1.61$, bounded by $[1.60,1.72]$.
These intervals express the amplitude resolution of the onset comparison.
At matched budget and source count, spatial rearrangement can advance or delay the response.

At fixed contrast budget, full support typically responds later than a single source across fifteen additional generator pairs, $G_5$-$G_{19}$.
These pairs sample different normalized directions in the same local $U(1)$-preserving operator basis.
Figure~\ref{Fig_15} compares their single-source and full-support responses at the same background amplitude and contrast budget as Fig.~\ref{Fig_14}.
At $L=14$, full support responds later in $26$ of $30$ pairs for $\Delta=0.6$ and $28$ of $30$ for $\Delta=1.8$.
Among pairs with both gap-ratio onsets observed, the median full-support-to-single-source ratios are $1.50$ and $1.57$, respectively.
The median full-support-to-single-source operator-scale ratios also exceed unity at $L=12$ for both probes and parents.

The same circuits can be expressed in terms of a local source amplitude $\eta$ multiplying $(s_b-\bar{s})\widehat{W}_c$.
For the odd-$K$ fields considered here, $\sum_b s_b^2=K$ and $\sum_b s_b=\pm1$, giving:
\begin{equation}
\epsilon=\eta\sqrt{K-\frac{1}{M}},
\quad
\frac{\eta_{r,50}^{\text{full}}}{\eta_{r,50}^{\text{single}}}
=\frac{1}{\sqrt{L}}
\frac{\epsilon_{r,50}^{\text{full}}}{\epsilon_{r,50}^{\text{single}}} \,.
\label{EqS_budget_local_amplitude}
\end{equation}
Panels~\ref{Fig_15}(e) and~\ref{Fig_15}(f) express the spectral comparisons in this local-amplitude convention.
At $L=14$, full support responds earlier in $59$ of $60$ comparisons across parents, generators, and starting signs, with one comparison unresolved.
The median ratios are $0.400$ and $0.419$ for $\Delta=0.6$ and $1.8$, respectively.
This conversion holds the background fixed; it is distinct from the original $\delta$ scan, which changes the common and contrast deformations together.
The factor $1/\sqrt{L}$ converts the full-support-to-single-source onset ratio from the budget convention to the local-amplitude convention.
It explains why full support typically responds later at equal budget but earlier at equal local amplitude.

For the stricter persistent joint spectral criterion, many fixed-budget cases do not reach the onset within $\epsilon\leq3$.
At $L=14$ and $\Delta=0.6$, full support responds earlier in $1$ comparison and later in $6$, with $23$ unresolved.
For $\Delta=1.8$, the corresponding counts are $5$, $17$ and $8$.

\begin{figure*}[htbp]
\centering
\includegraphics[width=6.7in]{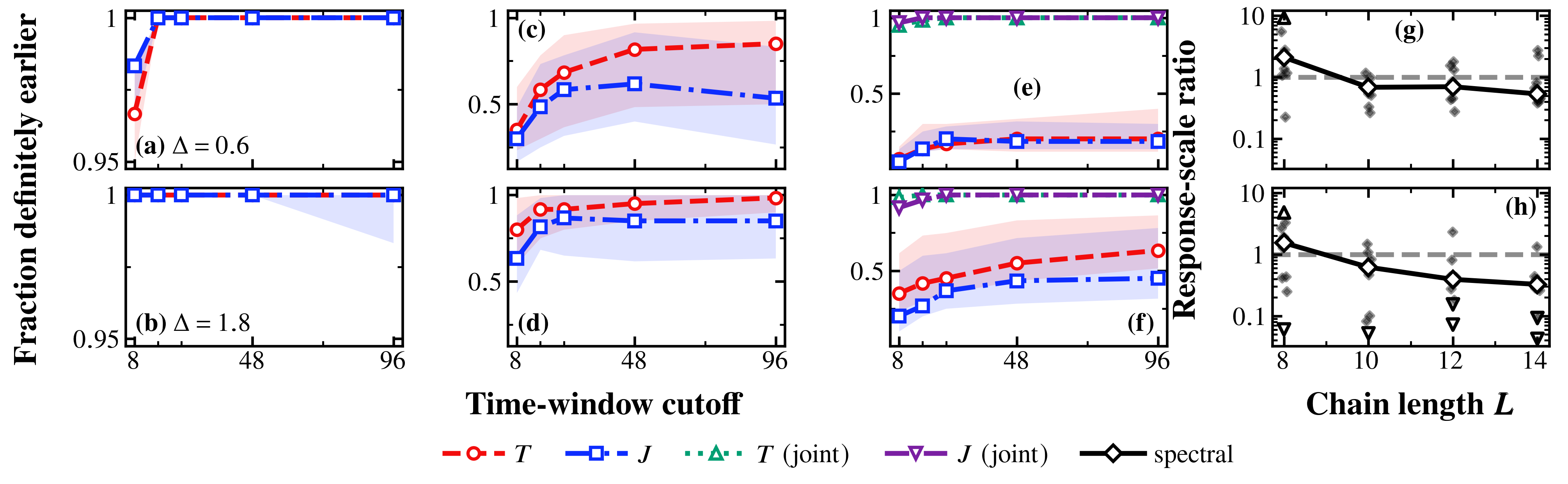}
\caption{{Sensitivity of the operator-spectrum ordering to thresholds and observation windows.
Upper and lower rows correspond to $\Delta=0.6$ and $\Delta=1.8$, respectively.
Panels (a) and (b) compare the absolute operator criterion with the joint spectral criterion; panels (c) and (d) use the gap-ratio criterion alone.
Red circles ($T$) and blue squares ($J$) use the original thresholds, and shading spans all $27$ threshold combinations.
For each parent and probe, the denominator includes all $60$ combinations of generator, system size, spatial profile, and starting sign, including unresolved cases.
Panels (e) and (f) use the relative $50\%$ operator criterion defined by Eq.~\eqref{EqS_relative_operator_response}.
Red and blue curves and shading compare with the three gap-ratio thresholds.
Green upward triangles ($T$) and purple downward triangles ($J$), labeled ``joint,'' compare with the joint criterion at its central threshold.
Panels (g) and (h) show the full-support-to-single-source ratio of $\delta_{r,50}$ with finer amplitude sampling than Fig.~\ref{Fig_12}; markers and bounds follow Fig.~\ref{Fig_14}.
One paired comparison at $L=10$ in panel (h) has no finite ratio bound and hence no plotted ordinate.
Shading describes threshold sensitivity, not statistical confidence.
Operator onsets generally precede the joint spectral onset near the original thresholds.
Using the gap-ratio criterion or relative operator response changes the fraction of cases in which the operator responds first.}}
\label{Fig_16}
\end{figure*}

Operator onsets remain below the persistent joint spectral onset over a range of thresholds and observation windows.
Figure~\ref{Fig_16} uses the original $\delta$ deformation and the expanded single-source and full-support fields with their sign-matched homogeneous references, rather than the fixed-$a$ budget comparison above.
The single source lies on the central bond for both starting signs.
We vary $t_{\text{max}}$ over $8$, $16$, $24$, $48$ and $96$ periods, and multiply the OSEE and OTOC thresholds independently by $0.75$, $1$ and $1.25$.
Together with gap-ratio envelope thresholds of $0.4$, $0.5$ and $0.6$, this gives $27$ threshold combinations.
The joint spectral condition retains the COE spacing preference, a rigidity gain of at least $0.05$, and persistence at two consecutive amplitudes.

Each threshold choice and time window gives $240$ operator-spectrum comparisons.
These cover five generators, two parents, three sizes, two spatial profiles, two starting signs and two probes.
At $t_{\text{max}}=16$, $24$ and $48$, every pair has an operator onset below the joint spectral onset for all $27$ threshold combinations.
At the original thresholds and $t_{\text{max}}=48$, both onsets are observed in $54$ pairs.
In the remaining $186$, the spectral onset lies beyond the scan range, but its lower bound still places it above the operator onset.
The largest upper bound on the operator-to-spectral onset ratio is $0.383$.

An exception occurs for $G_4$ at $\Delta=1.8$, $L=12$ and $t_{\text{max}}=96$, using the positive-sign single source and the $J$ probe.
Increasing the OTOC threshold by $25\%$ places the operator onset in $[0.78125,0.8125]$, after the joint spectral onset in $[0.575,0.600]$.
The operator onset remains later than the joint spectral onset for all three tested OSEE thresholds.
Using the gap-ratio criterion alone changes the ordering more substantially.
For example, consider the $J$ probe at the original operator thresholds, $t_{\text{max}}=48$, $\Delta=0.6$ and $L=12$.
The operator responds earlier in eight of twenty pairs and later in ten, with two unresolved.

To distinguish a small absolute response from an order-one fraction of the response across the deformation interval, we also form:
\begin{equation}
\begin{aligned}
D_X(\delta;t_{\text{max}}) ={}& \frac{1}{\sqrt{2}}\left[\left(\frac{\Delta S_{\mathrm{OSEE},X,\mathrm{RMS}}}{0.15\text{ bits}}\right)^2\right. \\
&\left. +\left(\frac{\Delta C_{\mathrm{OTOC},X,\mathrm{RMS}}}{0.04}\right)^2\right]^{1/2} \,.
\end{aligned}
\label{EqS_relative_operator_response}
\end{equation}
{The relative operator scale is the first amplitude at which the running envelope of $D_X$ reaches half its maximum over $0\leq\delta\leq1.3$.
This is a response fraction within the tested deformation interval, not an assumed asymptotic saturation value.
Panels~\ref{Fig_16}(e) and~\ref{Fig_16}(f) compare this relative operator onset with the gap-ratio and joint spectral criteria.
With finer amplitude sampling [Figs.~\ref{Fig_16}(g,h)], full support still has an earlier gap-ratio onset in seventeen of twenty pairs at $L=14$.
This includes eight of ten pairs at $\Delta=0.6$ and nine of ten at $\Delta=1.8$.}

To identify the origin of the fixed-budget response, we compare the alternating full-support field with rearrangements of exactly the same local gates in Eq.~\eqref{EqS_budget_gates}.
The background remains $a=0.04$, and the calculations use $L=12,14$, both parents, the twenty generator pairs $G_0$-$G_{19}$ and both contrast signs.
All nonuniform full-support arrangements have the same total numbers of the two gate types and the same summed squared local-gate distance from the homogeneous reference at every $\epsilon$.
In the block arrangement, the first $L/2$ bonds have $s_b=+1$, and the remaining bonds have $s_b=-1$.
Three scrambled arrangements preserve the positive and negative counts of this block field separately within each brickwall layer.

Different bonds contribute amplitudes to the same transition between reference eigenstates.
Let $U_H=U(a,0)$ and $K=\mathrm{i}(\partial_\epsilon U)U_H^\dagger|_{\epsilon=0}$.
We denote the unit-weight contribution of bond $b$ to the cyclically adjacent matrix element of $K$ by $\kappa_b(n,n+1)$.
This contribution includes conjugation by the remaining gates within the period.
The coherent transition weight and its self-only counterpart are:
\begin{equation}
T_n=\left|\sum_b q_b\kappa_b(n,n+1)\right|^2,
\quad
S_n=\sum_b q_b^2|\kappa_b(n,n+1)|^2 \,.
\label{EqS_channel_interference}
\end{equation}
{For nonvanishing $S_n$, a ratio $T_n/S_n<1$ identifies destructive interference in that channel.
For alternating full support, the typical ratios are $0.084$ and $0.116$ at $\Delta=0.6$ and $1.8$, respectively [Fig.~\ref{Fig_17}(c)].
They increase to $0.556$ and $0.603$ after scrambling, compared with $0.911$ and $0.936$ for the central source.
About $75\%$ and $73\%$ of the adjacent channels are destructive for the alternating field.
Some other channels are enhanced, so an average over matrix elements can obscure this suppression of typical nearby-level couplings.

\begin{figure*}[htbp]
\centering
\includegraphics[width=6.7in]{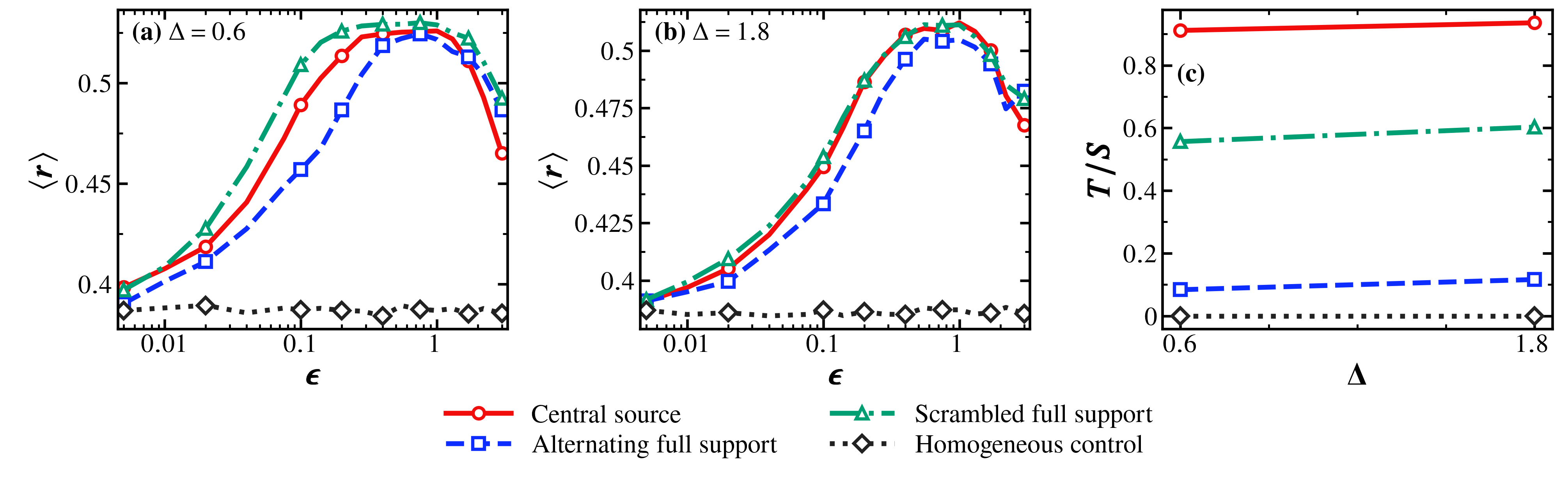}
\caption{{Spatial rearrangement and coherent interference at $L=14$ and $a=0.04$.
Panels (a,b) compare the gap ratio for the central source, alternating full support, scrambled full support and a homogeneous control.
The alternating and scrambled circuits use the same local gates and contrast budget; scrambling shifts the spectral crossover to smaller $\epsilon$.
Panel (c) shows the median of $T_n/S_n$ over adjacent reference levels, defined in Eq.~\eqref{EqS_channel_interference}, and reveals stronger typical cancellation for the alternating field.
Values are averaged over signs and, where applicable, three arrangements within each generator, then the median over twenty generators is shown.
The homogeneous control uses $q_b=\pm1/\sqrt{M}$ and has zero centered contrast budget; it remains integrable.
Lines connect all calculated points; markers identify selected points for readability.}}
\label{Fig_17}
\end{figure*}

The finite-amplitude spectra confirm the effect of rearrangement [Figs.~\ref{Fig_17}(a,b)].
At $L=14$, the median scrambled-to-alternating ratio of $\epsilon_{r,50}$ is $0.375$ for $\Delta=0.6$ and $0.614$ for $\Delta=1.8$.
For each generator, we first take the geometric mean of the ratios over signs and scrambled arrangements.
For each parent and each of the $40\%$, $50\%$ and $60\%$ gap-ratio thresholds, all twenty generator-level comparisons place the scrambled onset earlier, including the sampled crossing intervals.
With gate counts matched within each layer, the scrambled-to-block onset ratios are $0.642$ and $0.518$ at the $50\%$ threshold.
Scrambled arrangements respond earlier for every generator in this comparison as well.
Thus spatial rearrangement changes the response independently of both the total budget and the gate counts in each layer.

Rearrangement changes the coupling between nearby integrable eigenstates, consistent with the eigenstate-mixing description of finite-size chaos~\cite{Bulchandani2022}.
At fixed budget, the alternating pattern suppresses typical adjacent couplings through coherent cancellation.
At fixed local amplitude, the additional total strength from multiple bonds can outweigh that suppression, as quantified by Eq.~\eqref{EqS_budget_local_amplitude}.
The quantities here resolve typical adjacent channels at a common fixed background; they are distinct from the all-pair averages along the original deformation.

To examine the role of probe location, we move the single source through every bond while keeping the probe and OSEE partition at the center.
We repeat the calculation for both the $T$ and $J$ probes.
We use $a=0.04$, $L=10,12$, $\epsilon=0.02,0.1,0.4$, both parents, twenty generators and both signs, with evolution up to $96$ periods.
For a source at bond $b$, let $O_b(t)$ be the evolved probe and let $O_{\mathrm{comp}}(t)$ use the same uniform compensating background with that source removed.
Then:
\begin{equation}
\begin{aligned}
D_{\mathrm{loc}}(t) &= \frac{\norm{O_b(t)-O_{\mathrm{comp}}(t)}_F}{\norm{O(0)}_F}, \\
\mathcal{D}_X(t_{\max}) &= \left[\frac{1}{t_{\max}}\sum_{t=1}^{t_{\max}}\frac{\norm{O_X(t)-O_H(t)}_F^2}{\norm{O(0)}_F^2}\right]^{1/2} \,.
\end{aligned}
\label{EqS_local_operator_distances}
\end{equation}
Here $O_H(t)$ is the common reference at $\epsilon=0$, and $X$ denotes the spatial arrangement.
Unlike the combined OSEE-OTOC response in Eq.~\eqref{EqS_relative_operator_response}, these distances directly compare operators.
The single-source and compensated-reference circuits differ only at bond $b$, so $D_{\mathrm{loc}}$ vanishes until the probe's circuit causal cone reaches that bond.
In all $9600$ defect cases, the operator response begins at the first causal contact.

\begin{figure*}[htbp]
\centering
\includegraphics[width=6.7in]{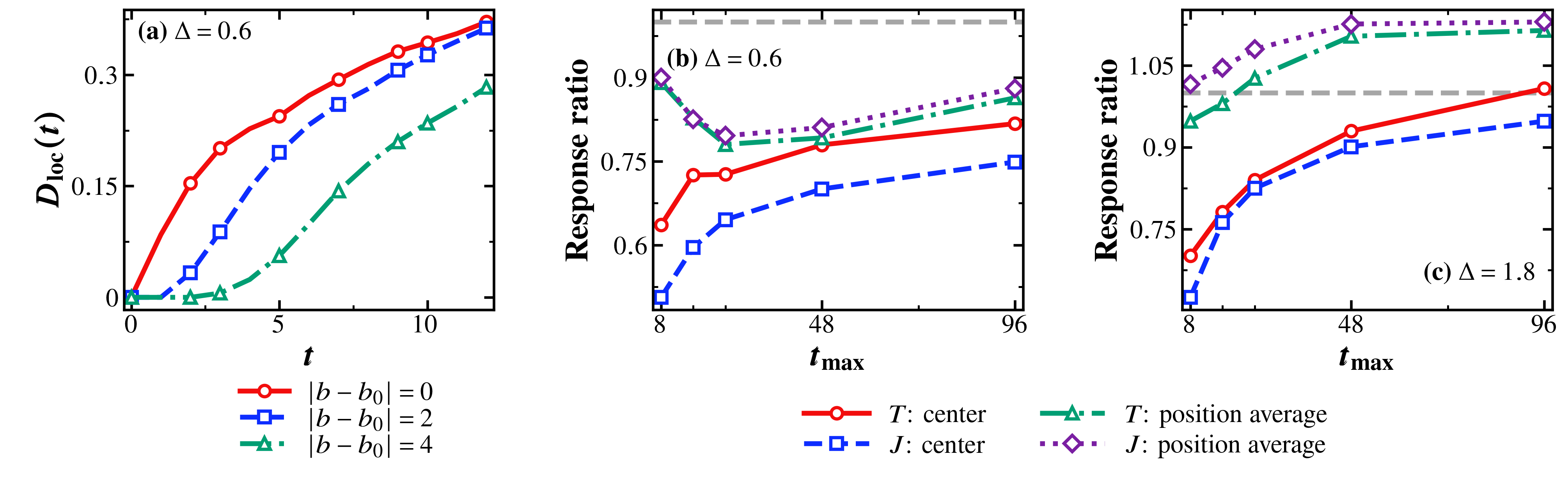}
\caption{{Probe location and observation-window dependence at $L=12$, $a=0.04$ and $\epsilon=0.1$.
Panel (a) shows $D_{\mathrm{loc}}(t)$ for the $T$ probe at $\Delta=0.6$, with the defect displaced from the central bond $b_0$ within the same brickwall layer.
More distant defects respond later because they enter the causal cone later.
Panels (b,c) show $\mathcal{D}_{\mathrm{full}}/\mathcal{D}_{b_0}$ (center) and $\mathcal{D}_{\mathrm{full}}/[M^{-1}\sum_b\mathcal{D}_b^2]^{1/2}$ (position average) for both probes at $\Delta=0.6$ and $1.8$.
Ratios and distances are averaged over the two signs within each generator before taking the median over twenty generators.
The gray line indicates equal response.
Position averaging reduces the early contrast with a central defect, while longer-window responses retain a dependence on the parent and probe.
Lines connect all sampled times or windows, with markers at selected points.}}
\label{Fig_18}
\end{figure*}

At $t_{\max}=8$, the full-support-to-central-source distance ratios range from $0.506$ to $0.701$ across the four combinations of parent and probe in Fig.~\ref{Fig_18}.
Using the position average instead gives $0.891$-$1.016$, showing that the central source's proximity contributes to its early response advantage.
At $t_{\max}=48$, the position-averaged ratios remain $0.792$-$0.811$ for $\Delta=0.6$ and $1.104$-$1.126$ for $\Delta=1.8$.
Propagation distance partly explains the early difference between full support and a central source.
At longer times, the response still depends on the parent circuit and probe.

Changes in propagation and phase accumulation can produce an operator response.
Applying the same gate deformation on every bond provides an integrable control~\cite{Znidaric2025a}.
At $L=12$, $\epsilon=0.1$ and $t_{\max}=48$, this control gives median operator distances of $0.49$-$0.63$ across the two parents and probes.
The corresponding OTOC RMS changes are $0.015$-$0.023$.
Thus an early operator response measures departure from the reference dynamics and need not coincide with the formation of chaotic spectral correlations.
\bibliographystyle{apsrev4-2}
\end{document}